\documentclass[11pt]{article}
\pdfoutput=1
\usepackage[centertags]{amsmath}
\usepackage[square, comma, sort&compress,numbers]{natbib}
\usepackage{array,multirow}
\numberwithin{equation}{section}
\usepackage{amssymb,amsfonts}
\usepackage{graphicx}
\usepackage{color}
\usepackage{mathtools,bm}
\usepackage{accents}
\usepackage{amsmath}
\usepackage{cancel}

\usepackage{epsfig}
\usepackage{bbold}
\usepackage{wrapfig}
\usepackage{float}
\usepackage{soul}
\usepackage{tkz-euclide}
\usepackage{braket}
\usepackage{tikz,pgf}
\usetikzlibrary{shapes}
\usetikzlibrary{calc}
\usetikzlibrary{decorations.pathmorphing}
\usetikzlibrary{decorations.pathreplacing,shapes.misc}
\usetikzlibrary{positioning}
\usetikzlibrary{arrows}
\usetikzlibrary{decorations.markings}
\usetikzlibrary{shadings}

\usetikzlibrary{intersections}

\newcommand{\be}{\begin{equation}}
\newcommand{\ee}{\end{equation}}
\newcommand{\ba}{\begin{array}}
\newcommand{\ea}{\end{array}}

\newcommand{\dps}{\displaystyle}
\newcommand{\half}{\frac{1}{2}}

\newcommand{\bref}[1]{\textbf{\ref{#1}}}

\newcommand{\im}{\mathop{\mathrm{Im}}}
\newcommand{\re}{\mathop{\mathrm{Re}}}

\newcommand{\CC}{\mathbb{C}}

\newcommand{\RR}{\mathbb{R}}

\newcommand{\ZZ}{\mathbb{Z}}
\newcommand{\NN}{\mathbb{N}}
\newcommand{\NO}{\mathbb{N}_0}

\newcommand{\cA}{\mathcal{A}}

\newcommand{\cC}{\mathcal{C}}
\newcommand{\cD}{\mathcal{D}}

\newcommand{\cF}{\mathcal{F}}

\newcommand{\cO}{\mathcal{O}}

\newcommand{\cV}{\mathcal{V}}

\numberwithin{equation}{section} \makeatletter
\@addtoreset{equation}{section}

\newcommand{\ads}{AdS$_2\;$}

\newcommand{\h}{h}
\newcommand{\cross}{\eta}

\newcommand{\dm}{(\delta {m}^2)}

\newcommand{\coef}{\kappa}

\newcommand{\sltwo}{sl(2,\mathbb{R})}

\newcommand{\bx}{{\bf x}}
\newcommand{\dc}{\cA^{\text{cont}}_4}
\newcommand{\de}{\cA^{\text{exch}}_4}
\newcommand{\DD}{\mathbb{H}} 
\newcommand{\PP}{\mathbb{P}}

\usepackage{jheppub}
\makeatletter
\def\@fpheader{\vspace{-.1cm}}
\makeatother

\title{\centering{Wilson network expansion  for four-point contact and exchange scalar Feynman diagrams in AdS$_2$}}

\dedicated{Dedicated to the memory of Igor V. Tyutin (1940--2026).}

\author{Konstantin\ Alkalaev}  
\author{and Vladimir\ Khiteev}

\affiliation{I.E. Tamm Department of Theoretical Physics, \\P.N. Lebedev Physical
Institute, 119991 Moscow, Russia}

\emailAdd{alkalaev@lpi.ru}
\emailAdd{khiteev@lpi.ru}

\abstract{We derive new  integral identities for AdS propagators and further develop  the Wilson network expansion for AdS Feynman diagrams.  In particular, we demonstrate  that four-point  contact and exchange  scalar diagrams in two dimensions can be expanded into several infinite series of matrix elements of Wilson line network operators with running conformal weights. Each series is  characterized by specific multi-trace operators associated with the external and intermediate edges of the corresponding  graphs. The resulting  expansions near the conformal boundary   reproduce the well-known decompositions of the corresponding four-point Witten diagrams into conformal blocks.}

\begin{document}

\maketitle
\flushbottom

\section{Introduction}

The study of Feynman diagrams on curved backgrounds is central to several areas of modern quantum field theory. On one hand, within the framework of holographic duality, these diagrams serve as a primary tool for perturbatively  calculating bulk correlation functions, which are essential for probing the local-scale dynamics of quantum gravity. On the other hand, exploring these correlators across diverse geometries uncovers universal features of quantum fields, offering a systematically controlled pathway toward understanding quantum gravity outside flat space. Furthermore, such investigations clarify how standard quantum field theory concepts, including vacuum definitions and particle states, adapt or break down in the absence of Poincaré invariance. This broader perspective ensures that insights gained from specific symmetric backgrounds can be properly generalized to dynamic, non-trivial spacetimes.

As a prime realization of this approach, anti-de Sitter (AdS) space serves as a highly controlled arena where the maximal isometry group constrains the algebraic structure of amplitudes and provides a natural infrared regulator (a radius $L_{\text{AdS}}$) for internal loops. However, even within this highly symmetric framework, the explicit calculation of Feynman diagrams remains a formidable challenge since this problem is further complicated by the fact that the bulk-to-bulk propagator is no longer a simple rational function as in the Minkowski case; instead, it is expressed in terms of the Gauss hypergeometric function. To circumvent the direct integration difficulties, we focus on decomposing these diagrams into a set of basis functions in the space of bulk correlators, in close analogy to the conformal block decomposition in conformal field theory (CFT). 

A central proposition of our previous studies \cite{Alkalaev:2025meb, Alkalaev:2024cje} is that these basis elements can be explicitly characterized as particular matrix elements of Wilson line networks extending into the AdS bulk. It was shown that these matrix elements are holographically reconstructed global conformal blocks. Consequently, such a decomposition allows one to reconstruct the AdS Feynman diagrams directly from the conformal blocks in the boundary CFT. This provides a constructive example of a bidirectional AdS/CFT correspondence: while boundary CFT correlation functions are obtained from the asymptotics of the bulk correlation functions, the inverse mapping reconstructs AdS correlation functions by decomposing them into bulk Feynman diagrams and expanding the latter in terms of conformal blocks via the Wilson network decomposition procedure. 

Thus far, our analysis has been limited to two- and three-point AdS Feynman diagrams of scalar fields in two-dimensional AdS space. Furthermore, motivated by the $1/N$ expansion, we have focused on these diagrams at the tree level.\footnote{From the boundary CFT perspective, the role of $N$ is played by the large central charge. Accordingly, we consider global conformal blocks corresponding to the $\sltwo$ subalgebra of the Virasoro algebra. In the bulk, $\sltwo$ acts as the isometry algebra of \ads space, where Wilson lines carry  its infinite-dimensional irreducible representations.}  In the present work, we take a further step by considering four-point \ads Feynman diagrams, which comprise contact and exchange contributions at tree level. As in our previous work, the decomposition procedure  consists of two parts: (i) establishing specific integral identities involving standard and modified AdS propagators, and  (ii) decomposing an AdS Feynman diagram by using these identities into infinite sums of  matrix elements of Wilson networks, which may possess different topologies than the original diagrams but share the same number of endpoints.  In this decomposition, the external and intermediate conformal weights are linear combinations of the original conformal weights  (masses), corresponding  to the weights  of single-trace, double-trace and, more generally, multi-trace primary operators in the boundary CFT.  
  
The paper is organized as follows. In Section \bref{sec:known_diagrams}, we review  the known expansions of a three-point AdS Feynman diagram in the bulk and four-point Witten diagrams on the boundary, while also establishing our conventions and notation. In Section \bref{sec:propagator_ids}, we formulate new integral identities for AdS propagators, relegating  their detailed derivations to Appendix \bref{app:derivation}. Along with the propagator identities established in \cite{Alkalaev:2025meb}, these new identities are required to expand four-point AdS diagrams. Section \bref{sec:decomp} presents the new integral representations of the matrix elements of the four-point Wilson line networks and the Wilson network expansions for four-point contact and exchange \ads Feynman diagrams: the relations \eqref{vert_decomp} and \eqref{exch_final}  along with their diagrammatic  representations in  figs.~\bref{fig:contact_vertex} and \bref{fig:ex_vertex} constitute the main result of this paper. The regularization procedure used to find the Wilson network expansions is discussed in detail in Appendix \bref{app:regularization}. Potential $n$-point generalizations of these results are shortly discussed in the concluding section \bref{sec:conclusion}.

\section{AdS diagram decompositions}
\label{sec:known_diagrams}

\subsection{Conformal block expansion of four-point Witten diagrams}


\begin{figure}
\centering
\includegraphics[scale=0.73]{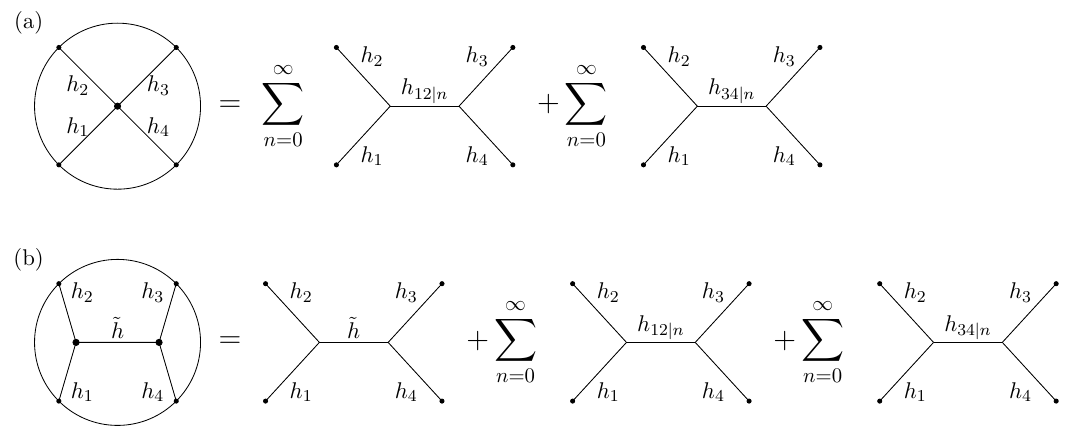}
\caption{Conformal block expansions of a four-point contact {\bf (a)} and exchange {\bf (b)} Witten diagrams.  On the left-hand side, four scalars with masses $m^2_i = h_i(h_i-1)$ are located in the boundary points $z_i$, $i=1,2,3,4$; central dots denote the AdS integrations;  the straight lines are  the standard bulk-to-bulk and bulk-to-boundary scalar propagators.  On the right-hand side:  each term is the 4-point  scalar conformal block with specified intermediate conformal weights  $\tilde h$ and $h_{ij|n} = h_i+\h_j+2n$.}
\label{fig:block_ex}
\end{figure}

Bulk correlation functions of the AdS fields can be decomposed into AdS Feynman diagrams comprising multiple integrals over the AdS space, where the integrands are products of bulk-to-bulk propagators. At the AdS boundary, such diagrams are conformally covariant and can be represented as an infinite  sum of conformal blocks. Four-point contact and exchange Witten diagrams in two dimensions have been extensively studied  \cite{Liu:1998th,Hoffmann:2000mx,Dolan:2001tt, Hijano:2015zsa} and take the form: 
\be 
\label{4pt_contact_expansion}
\ba{l}
\dps
\dc(z_i) = \int_{\text{AdS}_2}d^2\bx\sqrt{g(\bx)} \; K_{\h_1}(\bx,z_1) K_{\h_2}(\bx,z_2)K_{\h_3}(\bx,z_3)K_{\h_4}(\bx,z_4)  
\vspace{3mm}
\\
\dps
\hspace{16mm}
= \sum_{n=0}^{\infty}c^{^{(12|34)}}_n\, F_{h,h_{12|n}}(z_1,...,z_4) + \sum_{n=0}^{\infty}c^{^{(34|12)}}_n\, F_{h,h_{34|n}}(z_1,...,z_4)\,;
\ea
\ee 
\be 
\label{4pt_exchange_expansion}
\ba{l}
\dps
\de(z_i) =\iint_{\text{AdS}_2}d^2\bx\,d^2\bx'\sqrt{g(\bx)g(\bx')} \; K_{\h_1}(\bx,z_1) K_{\h_2}(\bx,z_2)G_{\tilde{h}}(\bx,\bx')K_{\h_3}(\bx',z_3)K_{\h_4}(\bx',z_4) 
\vspace{3mm}
\\
\dps
\hspace{0mm}= c_0\, F_{h,\tilde{h}}(z_1,...,z_4) + \sum_{n=0}^{\infty}\frac{c^{^{(12|34)}}_n}{\dm^{h_{12|n}}_{\tilde{h}}}\, F_{h,h_{12|n}}(z_1,...,z_4)+ \sum_{n=0}^{\infty}\frac{c^{^{(34|12)}}_n}{\dm^{h_{34|n}}_{\tilde{h}}}\, F_{h,h_{34|n}}(z_1,...,z_4)\,.
\ea
\ee 
The graphical representations of these expansions are shown in fig.~\bref{fig:block_ex}. The following terms appear in the relations above. 

\noindent {\bf 1.} The Poincare patch of the Euclidean AdS$_2$ in the local coordinates $\bx = (u,z)$ is given by the interval   
\be 
\label{metric}
ds^2 = \frac{dz^2+du^2}{u^2}\,,
\qquad
u\in\mathbb{R}_{\geq 0}\,,\; z \in \mathbb{R}\,.
\ee
The conformal boundary of  global \ads is disconnected; we choose the component located at $u=0$. The metric determinant $g(\bx) = u^{-4}$. 

\noindent {\bf 2.} $G_h(\bx,\bx')$ is the bulk-to-bulk propagator of a scalar field of mass $m^2 = h(h-1)$, $h\geq\half$ \cite{Fronsdal:1974ew}:
\be 
\label{bulk-to-bulk}
G_h(\bx,\bx') = \left(\frac{\xi(\bx,\bx')}{2}\right)^h{}_2F_1\left[\frac{h}{2},\frac{h}{2}+\half; h+\half\Big|\,\xi(\bx,\bx')^2\right],\quad \xi(\bx,\bx')= \frac{2u u'}{u^2+u'^2+(z-z')^2}\,;
\ee
the bulk-to-boundary propagator $K_h(\bx,z')$ is obtained by taking  one of the bulk points to the conformal boundary,  
\be 
\label{bulk-to-boundary}
K_h(\bx,z') = \left(\frac{u}{u^2 + (z-z')^2}\right)^h.
\ee 

\noindent {\bf 3.} Recall that Virasoro conformal blocks in CFT serve as kinematic basis functions for expanding conformal correlation functions. For instance, a four-point correlation function is decomposed as
\be 
\braket{O_1(z_1)O_2(z_2)O_3(z_3)O_4(z_4)} = \sum_{\tilde{h}}\cC_{h_1h_2\tilde{h}}\cC_{\tilde{h}h_3h_4}F_{h,\tilde{h},c}(z_1,...,z_4)\,,
\ee 
where $O_i(z_i)$ are primary operators with conformal weights $h_i$, the coefficients $\cC_{h_a h_b h_c}$ are model-dependent structure constants, the sum runs over the primary spectrum of the theory, and $c$ is the central charge. The global conformal blocks correspond to the $c\to\infty$ limit of the Virasoro blocks, with the weights scaled as $h_i, \tilde{h}= \cO(c^0)$. The functions $F_{h,h'}(z_1,...,z_4)$ in \eqref{4pt_contact_expansion} and \eqref{4pt_exchange_expansion} are the four-point global scalar conformal blocks with intermediate weights $h'=\tilde h$ or $h'=h_{ij|n} = h_i+h_j+2n$ and  external weights $\{h_1, ..., h_4\}$, which, by a slight abuse of notation, are collectively denoted as $h$.

\noindent {\bf 4.} The $c$-coefficients are given by 
\be 
\label{coef_4pt}
\ba{l}
\dps
c_0 =  \beta_{\tilde{h}\h_1\h_2}\beta_{\tilde{h}\h_3\h_4}\sum_{m,n=0}^{\infty}\frac{a^{h_1h_2}_m}{\dm ^{\tilde{h}}_{h_{12|m}}\,\beta_{h_{12|m}\h_1\h_2}}\,\frac{a^{h_3h_4}_n}{\dm ^{\tilde{h}}_{h_{34|n}}\,\beta_{h_{34|n}\h_3\h_4}}\,,
\vspace{3mm}
\\
\dps
c^{^{(ij|kl)}}_n = a^{h_ih_j}_n\, \beta_{h_{ij|n}\h_k\h_l}\,\sum_{m=0}^{\infty}\frac{a^{h_kh_l}_m}{\dm^{h_{ij|n}}_{h_{kl|m}}\,\beta_{h_{kl|m}\h_k\h_l}}\,,
\ea
\ee 
where 
\be
\label{a_beta}
\ba{c}
\dps
\beta_{h_1h_2h_3} = \frac{\Gamma(\frac{h_1+h_2-h_3}{2})\Gamma(\frac{h_1+h_3-h_2}{2})}{2\Gamma(h_1)}\,,
\qquad
a^{h_1h_2}_m = \frac{(-)^m(\h_1)_m(\h_2)_m}{m!(\h_1+\h_2-\half+m)_m}\,,
\vspace{3mm}
\\
\dps
\dm^{h_1}_{h_2} = \frac{\Gamma(h_2)}{2\pi^{\half}\Gamma(h_2+\half)}(h_1(h_1-1) - h_2(h_2-1))\;,
\ea
\ee
and $(a)_n = \Gamma(a+n)/\Gamma(a)$ is the Pochhammer symbol. By definition, $\dm^{h_1}_{h_2}$ is proportional to the difference between the two mass terms  corresponding  to the respective quadratic Casimir operator eigenvalues. 

\subsection{Wilson network expansion of  three-point AdS Feynman diagram}

The $n$-point Wilson line network \cite{Alkalaev:2023axo} (see also  \cite{Bhatta:2016hpz,Besken:2016ooo,Bhatta:2018gjb,Castro:2018srf,Alkalaev:2020yvq}) is built by contracting  an $n$-valent $\sltwo$ intertwiner carrying $n$ external and $n-3$ internal weights $h_1,...,h_n$ and $\tilde{h}_1,...,\tilde{h}_{n-3}$ with $n$ Wilson line operators of weights $h_1,..., h_n$ connecting  the point $\bx_0 = (1,0)$ to the points $\bx_1,..., \bx_n$ in the bulk. The resulting Wilson line network operator  acts on the tensor product of $n$ infinite-dimensional $\sltwo$ Verma modules. Let $\cV_{h,\tilde{h}}(\bx_1,..., \bx_n)$  be a  matrix element of this operator evaluated  between $n$ Ishibashi  states. In what follows, we refer to these matrix elements  as $n$-point AdS vertex functions. For example, the three-point AdS vertex function is defined as 
\be 
\label{3-p_def}
\cV_{ h_1h_2h_3}(\bx_1,\bx_2,\bx_3)= \braket{a_1|W_{\h_1}[\bx_0,\bx_1]I_{h_1h_2h_3}W_{h_2}[\bx_2,\bx_0]W_{h_3}[\bx_3,\bx_0]|a_2}\otimes\ket{a_3}\,,
\ee
where 
\begin{itemize}
\item The Ishibashi state $\ket{a_i}$ belongs to the infinite-dimensional $\sltwo$ Verma module $\cD_{h_i}$ with conformal  weight $h_i$. This module is described in the ladder basis:
\be 
\{\cD_h \ni  \ket{h,m}:\, J_0\ket{h,m} = m \ket{h,m},\, m = -h, -h-1, -h-2, ..., -\infty\}\,,
\ee 
where  the $\sltwo$ basis elements  $J_0, J_{\pm 1}$ satisfy the commutation relations $[J_k,J_l] = (k-l)J_{k+l}$ for $k,l = 0, \pm1$. The Ishibashi states are given by 
\be 
\ket{a_i} = \sum_{n=0}^\infty \frac{(-)^n\Gamma(2h_i)}{2^nn!\Gamma(2h_i+2n)} \, (J_1)^{2n}\ket{h_i,-h_i}\,.
\ee 

\item The Wilson line operator is defined as $W_h[\gamma] = \mathbb{P}e^{-\int_{\gamma}A}$, where $\gamma$ is a path from  $x_1$ to $x_2$ in AdS$_2$ and $A$ is the flat $\sltwo$ connection taking values in $\cD_h$.  Specifically, it is given by 
\be 
\label{Wilson_operator}
W_h[x_1,x_2]=u_2^{J_0}e^{z_{_{12}} J_1}u_1^{J_0}\,,
\ee 
where $z_{ij} = z_i-z_j$ \cite{Banados:1994tn}. Note that due to flatness, the Wilson line operator depends only on the endpoints of the path $\gamma$.

\item The $3$-valent intertwiner $I_{h_1h_2h_3}$ is an invariant operator that projects the tensor product of two $\sltwo$ modules onto a third one,  
\be 
\label{intertwiner}
I_{h_1 h_2 h_3}:\quad\cD_{h_2}\otimes\cD_{h_3}\xrightarrow{}\cD_{h_1}\,.
\ee
The matrix elements of this intertwiner are the Clebsch-Gordan coefficients.

\end{itemize}

The AdS vertex functions can be viewed as a convenient basis in the space of bulk correlation functions, enabling a systematic decomposition of AdS Feynman diagrams. In the three-point  case, this idea is realized by the following expansion \cite{Alkalaev:2025meb}:\footnote{Note that the definition of the AdS vertex function here differs from that in \cite{Alkalaev:2024cje}. For convenience, we have divided the original expression by  the coefficient $C_{h,\tilde{h}}$, the explicit form of which is given by eq.  $(2.18)$ in \cite{Alkalaev:2024cje}.}
\be 
\label{3pt_expansion}
\ba{l}
\dps
\cA^{\text{cont}}_3(\bx_i) =  \int_{\text{AdS}_2}d^2\bx\sqrt{g(\bx)}\, G_{\h_1}(\bx,\bx_1) G_{\h_2}(\bx,\bx_2)G_{\h_3}(\bx,\bx_3) 
\vspace{3mm}
\\
\dps
= \coef_0\,\cV_{\h_1\h_2\h_3}(\bx_1,\bx_2,\bx_3)
+\bigg\{\sum_{n=0}^{\infty}\coef^{^{(1|23)}}_n\cV_{h_{23|n} \h_2\h_3}(\bx_1,\bx_2,\bx_3) + (1 \leftrightarrow 2)+ (1 \leftrightarrow 3)\bigg\}\,,
\ea
\ee 
where the coefficients are given by  
\be 
\label{d_coef}
\ba{l}
\dps 
\coef_0 =  \frac{\pi^{\half}}{2}\Gamma\Big(\frac{\h_1+\h_2+\h_3}{2}-\half\Big)\, \frac{\Gamma(\frac{-\h_1+\h_2+\h_3}{2})\Gamma(\frac{\h_1-\h_2+\h_3}{2})\Gamma(\frac{\h_1+\h_2-\h_3}{2})}{\Gamma(\h_1)\Gamma(\h_2)\Gamma(\h_3)}\,,
\quad
\coef^{^{(i|jk)}}_n = \frac{a^{h_jh_k}_n}{\dm^{h_i}_{h_{jk|n}}}\,.
\ea
\ee

This decomposition is closely related to the conformal block expansion of the four-point Witten diagrams \eqref{4pt_contact_expansion}--\eqref{4pt_exchange_expansion}. Here, the three-point AdS vertex functions serve as bulk counterparts to the boundary conformal blocks (see our discussion in \cite{Alkalaev:2025meb}). This correspondence is expected, given that $n$-point AdS vertex functions can be identified with HKLL reconstructed $n$-point global conformal blocks \cite{Alkalaev:2024cje}\footnote{HKLL construction provides a way to reconstruct scalar fields in the bulk from the primary fields located at the boundary by integrating them with the smearing functions \eqref{smear} \cite{Hamilton:2005ju, Hamilton:2006az}. Here, we use the similar construction where we reconstruct AdS vertex functions in the bulk from the conformal blocks on the boundary, see \cite{Alkalaev:2024cje} for details.}
\be 
\label{HKLL_vertex}
\hspace{-2mm}\cV_{h,\tilde{h}}(\bx_1,...,\bx_n)=
 \prod_{i = 1}^n \int_{z_i-iu_i}^{z_i+iu_i}dw_i\;\mathbb{K}_{h_i}(\bx_i,w_i) F_{h, \tilde{h}}(w_1,...,w_n)\,,
\ee
where $F_{h, \tilde{h}}(w_1,...,w_n)$ is the $n$-point global conformal block in the comb channel
\be
\label{comb_fin}
\ba{l}
\dps
\cF_{\h, \tilde{\h}}(z_1,...,z_n)=\prod_{l=2}^{n-1}\Big(\frac{z_{_{l+1\,l-1}}}{z_{_{l+1\,l}}z_{_{l-1\,l}}}\Big)^{\h_l}\Big(\frac{z_{_{23}}}{z_{_{12}}z_{_{13}}}\Big)^{\h_1}\Big(\frac{z_{_{n-2\,n-1}}}{z_{_{n\,n-1}}z_{_{n\,n-2}}}\Big)^{\h_n}\prod_{k=1}^{n-3}\cross_k^{\tilde{\h}_k}
\vspace{2.5mm}  
\\
\dps
\times F_{K} \left[\begin{array}{cccc}
     \h_1+\tilde{\h}_1-\h_2,&\tilde{\h}_1+\tilde{\h}_2-\h_3,&...\,,&\h_n+\tilde{\h}_{n-3}-\h_{n-1} \\
     2\tilde{\h}_1, &2\tilde{\h}_2,&...\,, &2\tilde{\h}_{n-3} \end{array};\cross_1,...\,,\cross_{n-3}\right]\,,
\ea 
\ee
where the arguments are given by the cross-ratios
\be 
\label{cross-ratios}
\cross_i=\frac{z_{_{i\,i+1}}z_{_{i+2\,i+3}}}{z_{_{i\,i+2}}z_{_{i+1\,i+3}}}\,,
\qquad
z_{_{i\,j}} = u_i -u_j\,.
\ee
The function $F_{K}$ is the comb function which is a hypergeometric-type function  of $N$ variables  \cite{10.1007/BF02392525,Rosenhaus:2018zqn}:
\be 
\label{def_comb}
\ba{l}
F_K \left[\begin{array}{ccccc}
     a_1,&b_1,&...\,,&b_{N-1}, &a_2 \\
     &c_1,&...\,, &c_N \end{array};z_1,...\,,z_N\right] 
\vspace{2mm}
\\ 
\dps
= \sum_{m_1,...,m_N=0}^{\infty}\frac{(a_1)_{m_1}(b_1)_{m_1+m_2}\dots(b_{N-1})_{m_{N-1}+m_N}(a_2)_{m_N}}{(c_1)_{m_1}\dots(c_N)_{m_N}}\frac{z_1^{m_1}}{m_1!}\cdots \frac{z_n^{m_N}}{m_N!}\,;
\ea 
\ee 
the smearing function $\mathbb{K}_h(\bx,w)$, which is the bulk-to-boundary propagator \eqref{bulk-to-boundary} of dual conformal weight $1-h$ is defined as:
\be
\label{smear}
\mathbb{K}_h(\bx,w) = \frac{-2i}{4^{h}}\frac{\Gamma(2h)}{\Gamma(h)\Gamma(h)}\left(\frac{u}{u^2 + (z-w)^2}\right)^{1-h}\,.
\ee 
Note that in the cases $n=2,3$ the global conformal block in \eqref{HKLL_vertex} reduces to the two-point and three-point correlation functions in CFT$_1$, respectively (omitting the structure constant in the three-point case). For example, the three-point AdS vertex function takes the form
\be 
\cV_{h_1h_2h_3}(\bx_1,\bx_2,\bx_3) = \prod_{i = 1}^3 \int_{z_i-iu_i}^{z_i+iu_i}dw_i\;\mathbb{K}_{h_i}(\bx_i,w_i) \frac{1}{w_{12}^{h_1+h_2-h_3}w_{23}^{h_2+h_3-h_1}w_{13}^{h_1+h_3-h_2}}\,.
\ee 
It is important to emphasize  that the integral on the right-hand side of \eqref{HKLL_vertex} converges only if the external conformal weights satisfy 
\be
\label{hi_cons}
h_i > 0\,, 
\quad
i=1,...,n\,,
\ee 
whereas the intermediate weights are all real (see \cite{Alkalaev:2025meb,Alkalaev:2024cje} for more detail). However,  since  we use the HKLL reconstruction formula \eqref{HKLL_vertex} in the context of the AdS Feynman diagrams, throughout the paper we impose additional restrictions on conformal weights:
\be
\label{half_restrictions}
\dps
\tilde{h} \geq \half\,,
\qquad
h_i \geq \half\,, 
\qquad
i=1,...,n\,,
\ee
which stem from the requirement  $h\geq\half$ in the definitions of the bulk-to-bulk \eqref{bulk-to-bulk} and bulk-to-boundary propagators \eqref{bulk-to-boundary}.

To further develop the use of AdS vertex functions as a basis for bulk correlation functions, we consider four-point contact and exchange AdS Feynman diagrams. Specifically, we show that these diagrams can be expanded into infinite series of AdS vertex functions and demonstrate that their  boundary asymptotics correctly recover the well-known conformal block expansions  \eqref{4pt_contact_expansion}--\eqref{4pt_exchange_expansion}.

\section{AdS propagator integral identities}
\label{sec:propagator_ids}

The decompositions described in the previous section were obtained by applying specific propagator identities to the bulk-to-bulk and bulk-to-boundary propagators within AdS Feynman diagrams. In this section, we provide an extended set of these identities, which will subsequently be used to decompose four-point diagrams in the bulk.  We begin by reviewing the three propagator identities from \cite{Alkalaev:2025meb} employed in the derivation of the three-point decomposition \eqref{3pt_expansion}, followed by the introduction of two additional identities. The complete set is summarized below. 

Note that, in general, the propagators are distributions, as they arise as solutions to differential equations involving distributional sources. Any propagator identity is therefore understood as an integral identity, where each distribution is integrated against a test function. For brevity, we sometimes omit the integrals and test functions in the relations  below.

\begin{figure}
\centering
\includegraphics[scale=0.9]{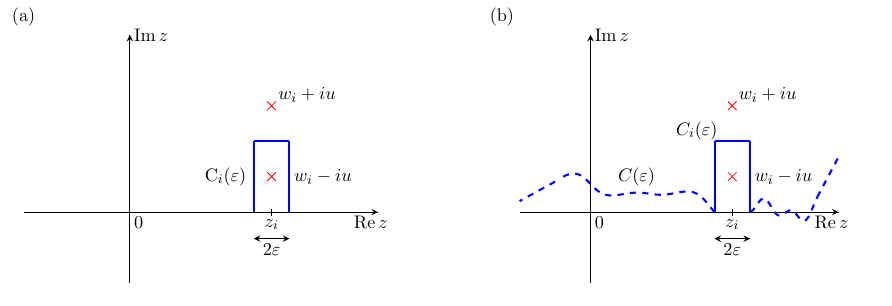}
\caption{{\bf (a)} The contours $C_i(\varepsilon)$ in the complex $z$-plane, $i=1,...,4$. The red crosses denote the poles of the modified propagators $\widehat{G}(\bx,\bx_i,w_i)$. {\bf (b)} An example of the contour $C(\varepsilon)$ (dashed line) in the complex $z$-plane. Note that the total  contour $C(\varepsilon) + C_i(\varepsilon)$ is always open due to the constraint on the contour $C(\varepsilon)$, namely $\re(z)\neq z_i$ for all $z\in C(\varepsilon)$. The contour $C(\varepsilon)$ does not cross the poles and branch cuts of the test function $f(\bx)$.}
\label{fig:C_i}
\end{figure}

\begin{itemize}
\item The conversion identity: 
\be 
\label{conversion_id}
G_h(\bx,\bx') = \int_{z'-iu'}^{z'+iu'}d w\;\widehat{G}_h(\bx,\bx',w)\,,
\qquad
\re(z)\neq \re(z')\;\; \text{or} \;\; u>u'-|\im(z-z')|\,,
\ee 
for  the modified propagator $\widehat{G}_h(\bx,\bx',w)$  defined by 
\be 
\label{G_hat}
\widehat{G}_h(\bx,\bx',w) = K_h(\bx, w)\mathbb{K}_{h}(\bx', w)\,.
\ee 
\item The superposition  identity: 
\be 
\label{superposition_id}
\ba{l}
\dps
\lim_{\varepsilon\to0}\int_{0}^{\infty} \frac{du}{u^2}\int_{C(\varepsilon)}dz\;G_{\h_i}(\bx,\bx_i)f(\bx) = \pi\int_{0}^{u_i} \frac{du}{u^2}\int_{z_i-i(u-u_i)}^{z_i+i(u-u_i)}dz\;\widetilde{G}_{h_i}(\bx,\bx_i)f(\bx)
\vspace{2.5mm}  
\\
\dps
\hspace{32mm}+\lim_{\varepsilon\to0}\int_{z_i-iu_i}^{z_i+iu_i}dw_i\int_{0}^{\infty} \frac{du}{u^2}\int_{C(\varepsilon)+C_i(\varepsilon)}dz\;\widehat{G}_{h_i}(\bx,\bx_i,w_i)f(\bx)\,,
\ea
\ee 
where the contour $C_i(\varepsilon)$ is shown in fig.~\bref{fig:C_i} {\bf (a)}; $C(\varepsilon)$ is a contour on the complex $z$-plane such that $C(\varepsilon)+C_i(\varepsilon)$ is a continuous curve; $C(\varepsilon)$ satisfies $\re(z)\neq z_i$ for all $z\in C(\varepsilon)$, see fig.~\bref{fig:C_i} {\bf (b)} for the example of such contour; $f(\bx)$ is a test function holomorphic on a domain $U \times V \subset \CC^2$, such that $\RR_+ \subset U $ and $C(\varepsilon)+C_i(\varepsilon) \subset V$; the modified propagator $\widetilde{G}_h(\bx,\bx')$ is given by
\be 
\label{G_tilde}
\widetilde{G}_h(\bx,\bx') = \frac{-2i}{4^{h}}\frac{\Gamma(2h)}{\Gamma(h)\Gamma(h)}\left(\frac{\Gamma(1-2h)}{\Gamma(1-h)^2}\, G_h(\bx,\bx') + \frac{\Gamma(2h-1)}{\Gamma(h)^2}\, G_{1-h}(\bx,\bx')\right).
\ee
\item The splitting identity:
\be 
\label{geodesic_split}
\ba{l}
\dps
\pi \int_{0}^{u_3} \frac{du}{u^2} \int_{z_3-i(u-u_3)}^{z_3+i(u-u_3)}\hspace{-2mm}dz\; G_{\h_1}(\bx,\bx_1) G_{\h_2}(\bx,\bx_2) \widetilde{G}_{h_3}(\bx,\bx_3) = \sum_{n=0}^{\infty}\frac{a^{h_1h_2}_n\,\alpha_{\h_1\h_2,n}}{\dm^{h_{12|n}}_{h_3}}
\vspace{2.5mm}  
\\
\dps
\times\int_{z_3-iu_3}^{z_3+iu_3} \hspace{-2mm} dw\;\oint_{0} \frac{du}{u^2} \oint_{P[w-iu,w+iu]} \hspace{-2mm} dz\, 
G_{\h_1}(\bx,\bx_1)G_{\h_2}(\bx,\bx_2) \widehat{G}_{h_{12|n}}(\bx,\bx_3,w)
\vspace{3mm}
\\
\dps
= \sum_{n=0}^{\infty}  \frac{a^{h_1h_2}_n}{\dm^{h_{12|n}}_{h_3}} \cV_{\h_1\h_2 \h_{12|n} }(\bx_1,\bx_2,\bx_3)\,,
\ea
\ee
where $P[w-iu,w+iu]$ is the Pochhammer contour and the $\alpha$-coefficient is given by
\be 
\label{alpha_prime}
\alpha_{h_1h_2,n} = \frac{(-)^{h_1+h_2+n}}{4\pi^{\frac{3}{2}}\sin(2\pi (h_1+h_2))}\frac{n!\,\Gamma(h_{12|n})}{\Gamma(h_1+h_2+n-\half)(h_1)_n(h_2)_n}\,.
\ee 
Note that any bulk-to-bulk propagator $G_{h_i}(\bx,\bx_i)$ in the first and  second lines of \eqref{geodesic_split} can be converted to the integral of the modified propagator $\widehat{G}_{h_i}(\bx,\bx_i,w_i)$ using the conversion identity \eqref{conversion_id}.

\end{itemize}

\noindent The following two identities provide a bulk generalization of the identities previously applied to four-point Witten diagrams \cite{Hijano:2015zsa}.

\begin{itemize}
\item The geodesic decomposition identity:
\be 
\label{geodesic_prop}
\ba{l}
\dps
\prod_{j=1}^2\int_{z_j-iu_j}^{z_j+iu_j}dw_j\; \widehat{G}_{h_1}(\bx,\bx_1,w_1)\widehat{G}_{h_2}(\bx,\bx_2,w_2) = \sum_{n=0}^\infty \frac{a^{h_1h_2}_n}{\beta_{h_{12|n}h_1h_2}} 
\vspace{3mm}
\\
\dps
\times \prod_{j=1}^2\int_{z_j-iu_j}^{z_j+iu_j}dw_j\;\int_{\gamma_{12}}d\lambda\, \widehat{G}_{h_1}(\bx(\lambda),\bx_1,w_1)\widehat{G}_{h_2}(\bx(\lambda),\bx_2,w_2)G_{h_{12|n}}(\bx(\lambda),\bx)\,,
\ea
\ee
where $|\xi(\bx,\bx(\lambda))|<1$ for all  $\lambda \in \RR$, $\xi(\bx,\bx(\lambda))$  is a geodesic distance \eqref{bulk-to-bulk} and the analytically continued geodesic $\bx(\lambda)\in\gamma_{12}$, $\bx(\lambda) = (u(\lambda),z(\lambda))$ connecting two points $w_1\in\CC$ and $w_2\in\CC$ is given by 
\be 
\label{comp_geod}
u(\lambda)^2 = \frac{(w_1-w_2)^2}{4\cosh^2\lambda}\,,
\qquad
z(\lambda) = \frac{(w_1+w_2)}{2} + \frac{(w_1-w_2)}{2}\tanh\lambda\,,
\ee 
see Appendix \bref{app:geod_dec} for the proof. 

This identity is the bulk version of the identity \eqref{bndry_geodesic_prop} from \cite{Hijano:2015zsa} which transforms the product of two bulk-to-boundary propagators into an infinite sum of geodesic integrals of two bulk-to-boundary and one bulk-to-bulk propagators. It is worth noting that the use of a geodesic as the integration contour is not strictly necessary for this identity to hold; the contour can be deformed to other paths in $\CC\times\CC$ that do not satisfy the geodesic equation.

\item The transition identity:
\be 
\label{transition_id}
\int_{C_u}\frac{du}{u^2}\int_{C_z}dz\; G_{h_1}(\bx,\bx_1)G_{h_2}(\bx,\bx_2)= \frac{1}{{\dm^{h_1}_{h_2}}}G_{h_1}(\bx_1,\bx_2)+\frac{1}{{\dm^{h_2}_{h_1}}}G_{h_2}(\bx_1,\bx_2)\,,
\ee
where $\bx_1,\bx_2\in\CC\times\CC$ and $|z_1-z_2|\geq|u_1+u_2|$; the integration contours $C_u$ and $C_z$ are shown in fig.~\bref{fig:C_u}. See Appendix \bref{app:transition} for the proof. This identity is an analytic continuation of  identity \eqref{transition_id_real} from \cite{Hijano:2015zsa}; it reduces the product of two bulk-to-bulk propagators with a common point $\bx$, integrated over  AdS$_2$, to a sum of two bulk-to-bulk propagators.  
\end{itemize} 

\begin{figure}
\centering
\includegraphics[scale=0.9]{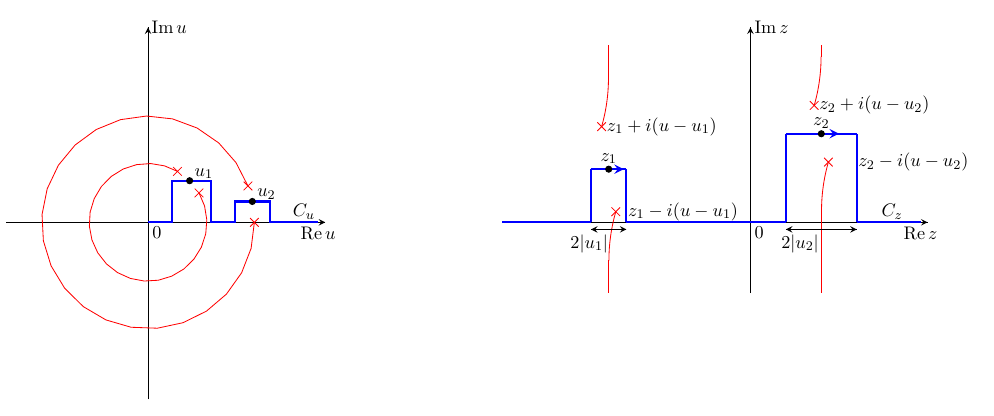}
\caption{Integration contours $C_u$ and $C_z$ in the complex $u$- and $z$-planes, respectively. Red crosses and lines denote the branch points and cuts of the integrand in \eqref{transition_id}.}
\label{fig:C_u}
\end{figure}

\section{Wilson  network representation for four-point AdS$_2$ Feynman diagrams}
\label{sec:decomp}

In this section, we explicitly derive the Wilson network expansion for two types of \ads Feynman diagrams: contact, 
\be
\label{4pt_diagram_cont_def}
\dc(\bx_i)=\int_{\text{AdS}_2} d^2\bx\;\sqrt{g(\bx)}\; G_{h_1}(\bx,\bx_1)G_{h_2}(\bx,\bx_2)
G_{h_3}(\bx,\bx_3)G_{h_4}(\bx,\bx_4)\,, 
\ee 
where the bulk-to-bulk propagators are given by \eqref{bulk-to-bulk} and the  triangle inequalities are required for the integral to converge:
\be
\label{triangle_identity1}
h_1 + \dots - h_k + \dots + h_4 > 0\,, 
\qquad 
\forall k \in \{1, ...\,, 4\}\,;
\ee
and exchange,
\be
\label{4pt_diagram_ex_def}
\de(\bx_i)=\iint_{\text{AdS}_2} d^2\bx \,d^2\bx'\sqrt{g(\bx)g(\bx')}\; G_{h_1}(\bx,\bx_1)G_{h_2}(\bx,\bx_2)G_{\tilde{h}}(\bx,\bx')G_{h_3}(\bx',\bx_3)  G_{h_4}(\bx',\bx_4)\,,
\ee
where  convergence  is ensured by the triangle inequalities: 
\be 
\label{triangle_identity2}
\ba{lll}
\dps h_i+h_j - h_k > 0  \,, & \qquad i\neq j\neq k\neq i\,, & \qquad i,j,k \in \{1,2,\sim\}\,,
\vspace{3mm}
\\
\dps h_n+h_m - h_p > 0 \,, & \qquad n\neq m\neq p\neq n\,, & \qquad n,m,p \in \{3,4,\sim\}\,,
\ea 
\ee
with the symbol  $\sim$ in the index sets denoting the intermediate weight $h_{\sim} \equiv \tilde h$.

\subsection{Integral representations of four-point AdS vertex functions}
\label{subsec:decom_cases}

The identities presented in the previous section allow one to decompose four-point AdS Feynman diagrams into an infinite sum of various integrals. To convert these integrals into four-point AdS vertex functions, we employ the HKLL representation \eqref{HKLL_vertex}: 
\be 
\label{4pt_HKLL}
\cV_{\h_1 \h_2 \h_3 \h_4,h}(\bx_1,\bx_2,\bx_3,\bx_4)=
 \prod_{i = 1}^4 \int_{z_i-iu_i}^{z_i+iu_i}dw_i\;\mathbb{K}_{h_i}(\bx_i,w_i) F_{\h_1 \h_2 \h_3 \h_4,h}(w_1,...,w_n)\,,
\ee
by substituting the four-point global conformal block $F_{\h_1 \h_2 \h_3 \h_4,h}(w_1,...,w_n)$ with one of its integral representations. Here, the external and internal conformal weights are $\{h_1,..., h_4\}$ and $h$, respectively. Although \eqref{4pt_HKLL} holds for any $h_i > 0$, $h\in\RR$  \eqref{hi_cons},  the use of integral representations of the conformal blocks and bulk-to-bulk, bulk-to-boundary propagators impose  additional restrictions on the weights (see our discussion below \eqref{hi_cons} and Appendix \bref{app:integral_vertex}). These restrictions partition the domain of possible conformal weights,  
\be 
\DD^+ = \left\{\bm h = (h_1,h_2,h_3,h_4,h)\in \RR^5\mid h_i, h\geq\half\right\}\,,
\ee 
which is an open half-space, into several regions. Specifically, we define:
\be
\label{total}
\DD^+ = \DD_{12}\cup \DD_1\cup \DD_2 =  \DD_{34}\cup\DD_3\cup \DD_4\,,
\ee
where 
\be 
\label{subdom}
\ba{c}
\dps
\DD_{12} = \{\bm h\in \DD^+: h_1<h+h_2,\, h_2<h+h_1\}\,,
\vspace{3mm}
\\
\dps
\DD_{34} = \{\bm h\in \DD^+: h_3 < h+h_4,\, h_4<h+h_3\}\,,
\vspace{3mm}
\\
\dps
\DD_1 = \{\bm h\in \DD^+: h_1\geq h+h_2\}\,,
\quad 
\DD_2 = \{\bm h\in \DD^+: h_2\geq h+h_1\}\,,
\vspace{3mm}
\\
\dps
\DD_3 = \{\bm h\in \DD^+: h_3\geq h+h_4\}\,,
\quad 
\DD_4 = \{\bm h\in \DD^+: h_4\geq h+h_3\}\,.
\ea
\ee 
Furthermore, each domain $\DD_i$ contains a family of equidistant planes, $\PP_i \subset \DD_i$:   
\be 
\label{subdomP}
\ba{c}
\dps
\PP_1 = \{\bm h\in \DD_1: h_1 = h_{h2|n},\, n\in\NO\}\,,
\quad 
\PP_2 = \{\bm h\in \DD_2: h_2 = h_{h1|n},\, n\in\NO\}\,,
\vspace{3mm}
\\
\dps
\PP_3 = \{\bm h\in \DD_3: h_3 = h_{h4|n},\, n\in\NO\}\,,
\quad 
\PP_4 = \{\bm h\in \DD_4: h_4 = h_{h3|n},\, n\in\NO\}\,,
\ea
\ee 
where $h_{hj|n} = h+h_j+2n$. These subsets are introduced because the Wilson network expansions necessarily generate terms with ``double-trace'' conformal weights via the splitting \eqref{geodesic_split} and geodesic decomposition \eqref{geodesic_prop} identities.  When a propagator with such a weight connects an endpoint to an internal vertex, the weights ${\bm h}$ of the resulting term fall into one of the discrete families of planes   $\PP_i$, requiring specific integral representations to complete the expansion.

Now, let us list the integral representations of the four-point AdS vertex functions corresponding to the subdomains \eqref{subdom} and \eqref{subdomP}. The proofs are provided  in Appendix \bref{app:integral_vertex}.

\begin{itemize}
\item The domain $\DD_{34}$:
\be 
\label{alt_geodesic_rep}
\ba{c}
\dps
\cV_{\h_1 \h_2 \h_3 \h_4,h}(\bx_1,\bx_2,\bx_3,\bx_4) = \frac{1}{\beta_{h h_3 h_4}}\prod_{j=3}^4\int_{z_j-iu_j}^{z_j+iu_j}dw_j \int_{\gamma_{34}}d\lambda'\, \cV_{\h_1 \h_2 h}(\bx_1,\bx_2, \bx(\lambda'))
\vspace{3mm}
\\
\dps
\times\, \widehat{G}_{h_3}(\bx(\lambda'),\bx_3,w_3)\,\widehat{G}_{h_4}(\bx(\lambda'),\bx_4,w_4)\,,
\ea
\ee 
where $\beta_{h h_3 h_4}$ is given by \eqref{coef_4pt}, $\cV_{\h_1 \h_2 h}(\bx_1,\bx_2, \bx(\lambda'))$ is the three-point AdS vertex function.

The integral representation in the domain $\DD_{12}$  is obtained from   \eqref{alt_geodesic_rep} by swapping the  indices of the conformal weights and coordinates as $\{1,2\} \leftrightarrow \{3,4\}$.

\item The domain $\DD_4 = \PP_{4} \cup (\DD_4 \setminus  \PP_4)$. Within $\PP_4$ we have
\be 
\label{non_triangle_rep}
\ba{c}
\dps 
\cV_{\h_1 \h_2 \h_3 \h_4,h}(\bx_1,\bx_2,\bx_3,\bx_4) = \alpha_{\h\h_3,n} \prod_{j=3}^4\int_{z_j-iu_j}^{z_j+iu_j}dw_j \; \oint_{0} \frac{du}{u^2} \oint_{P[w_4-iu,w_4+iu]} \hspace{-2mm} dz
\vspace{3mm}
\\
\dps
\hspace{15mm}\times\,\cV_{\h_1 \h_2 h}(\bx_1,\bx_2, \bx) \,\widehat{G}_{h_3}(\bx,\bx_3,w_3)\,\widehat{G}_{h_4}(\bx,\bx_4,w_4)\,,
\ea
\ee
where the $\alpha$-coefficient is given by \eqref{alpha_prime}. In the complement $\DD_4 \setminus  \PP_4$ we employ the same representation  \eqref{non_triangle_rep}, but with the $u$-integration contour replaced by  the Hankel contour to avoid branch cuts of the integrand.\footnote{This integral representation can be analytically continued to  $\DD_{12}$ or  $\DD_{34}$. By construction, the resulting  representation is equivalent to the integral representation \eqref{alt_geodesic_rep} which is more convenient in our analysis.}

\item Other domains. Similar representations  in  the domains  $\PP_{1},\PP_2,\PP_3$ are obtained  by relabelling the conformal weights and coordinates in \eqref{non_triangle_rep}. Specifically, the condition $h_4 = h_{h3|n}$ is replaced by $h_i=h_{hj|n}$ via the exchange $4\leftrightarrow i$ and $3\leftrightarrow j$. For instance,  the representation for  $h_2 = h_{h1|n}$ is obtained by  transposing   $4\leftrightarrow 2$ and $3\leftrightarrow 1$. 

For the remaining domains  $\DD_1 \setminus  \PP_1$, $\DD_2 \setminus  \PP_2$, $\DD_3 \setminus  \PP_3$ other integral representations are modified by the same relabelling of indices. These cases are not used in the expansion of the four-point AdS Feynman diagrams. 
   
\end{itemize}

Another useful representation of the four-point AdS vertex function is derived from the geodesic Witten diagram representation of the four-point global conformal block \cite{Hijano:2015zsa}, valid  in the domain $\DD_{12} \cap\DD_{34}$. Using the HKLL representation \eqref{4pt_HKLL}, we obtain
\be 
\label{geodesic_rep}
\ba{c}
\dps
\cV_{\h_1 \h_2 \h_3 \h_4,h}(\bx_1,\bx_2,\bx_3,\bx_4) = \frac{1}{\beta_{h h_1 h_2}\beta_{h h_3 h_4}}\prod_{j=1}^4\int_{z_j-iu_j}^{z_j+iu_j}dw_j \int_{\gamma_{12}}d\lambda\;\int_{\gamma_{34}}d\lambda'\, \widehat{G}_{h_1}(\bx(\lambda),\bx_1,w_1)
\vspace{3mm}
\\
\dps
\times\, \widehat{G}_{h_2}(\bx(\lambda),\bx_2,w_2)\,G_{h}(\bx(\lambda),\bx(\lambda'))\,\widehat{G}_{h_3}(\bx(\lambda'),\bx_3,w_3)\,\widehat{G}_{h_4}(\bx(\lambda'),\bx_4,w_4)\,,
\ea
\ee 
where the geodesics $\bx(\lambda)\in\gamma_{12}$ and  $\bx(\lambda')\in\gamma_{34}$ connect the boundary points $z_1,z_2$ and $z_3,z_4$ respectively, see \eqref{comp_geod}.\footnote{Note that the subdomains defined in \eqref{subdom} have non-empty overlaps, meaning that the conformal block in \eqref{4pt_HKLL} admits multiple equivalent integral representations for certain sets of weights.}

\subsection{Decomposition algorithm}
\label{sec:algorithm}

\noindent The general decomposition algorithm consists of the following four steps:

\begin{enumerate}

\item[I.]  Apply the superposition identity \eqref{superposition_id} to any bulk-to-bulk propagator $G$ connecting an endpoint to an internal vertex point of the AdS Feynman diagram. This splits the integral into a  sum of two terms: $\int GG... \to \int \widehat{G}G... + \int \widetilde{G}G...\;$. Repeat this process for each resulting term as long as the superposition identity remains applicable. For example, the identity cannot be applied to $G$ in the second term $\widetilde{G}G...$ due to a mismatch between its integration domain and that of $\widetilde{G}G$ (compact  vs. non-compact domains). In cases where  $G$ connects an endpoint to an internal vertex but  the superposition identity is inapplicable, use the conversion identity \eqref{conversion_id} to transform $G$ into $\widehat{G}$. Upon completion, no term should contain  a propagator $G$ connecting an endpoint and an internal vertex.

\item[II.] Apply the geodesic decomposition identity \eqref{geodesic_prop} to every product of two modified propagators $\widehat{G}_{h_i}(\bx,\bx_i,w_i)\widehat{G}_{h_j}(\bx,\bx_j,w_j)$ that is integrated over non-compact domain in $\bx$. 

\item[III.] Apply the transition identity \eqref{transition_id} wherever possible to further simplify the expressions and then convert each term involving $\widetilde{G}$ into an infinite sum of terms containing  three-point AdS vertex functions using  the splitting identity \eqref{geodesic_split}. After this step, all non-compact AdS$_2$ integrations should be resolved, yielding  infinite sums of geodesic Feynman diagrams or integrals of the form \eqref{alt_geodesic_rep}, \eqref{non_triangle_rep}.

\item[IV.] Replace each term in the resulting sum with the corresponding AdS vertex function using the integral representations \eqref{alt_geodesic_rep}, \eqref{non_triangle_rep}, or \eqref{geodesic_rep}. 

\end{enumerate}

It is assumed that at each step of the algorithm, the integration contours do not cross any branch cuts of the integrand. Before applying any identity, its domain of applicability (i.e., restrictions on conformal weights and coordinates) must be verified. If an identity cannot be applied because a contour lies outside this domain, the contour should be deformed to ensure the identity's validity.

\subsection{Four-point contact diagram}
\label{sec:contact}

Applying the above algorithm to a four-point contact AdS Feynman diagram \eqref{4pt_diagram_cont_def} yields the following scheme: 
$$
\ba{rcl}
\dps \int_{\text{AdS}_2} G_1 G_2 G_3 G_4 
& \xrightarrow[\hspace{6.3mm}\rm (I)\hspace{6.3mm}]{\eqref{superposition_id},\, \eqref{conversion_id}} & 
\dps 
\ba{l} 
\dps
\hspace{3.5mm}\int_{\text{AdS}_2} \widehat{G}_1\widehat{G}_2\widehat{G}_3\widehat{G}_4 
+ \int_{\text{AdS}_2} \widetilde{G}_1\widehat{G}_2\widehat{G}_3\widehat{G}_4 
+ (1 \leftrightarrow 2) + (1 \leftrightarrow 3) + (1 \leftrightarrow 4) 
\ea

\ea
$$

\be 
\label{scheme}
\ba{rcl}
\dps

& \xrightarrow[\hspace{6mm}\rm (II)\hspace{6mm}]{\eqref{geodesic_prop}} & 
\dps 
\ba{l} 
\dps
\hspace{3.5mm}\sum_{n,m}\beta^{12}_m\beta^{34}_n \int_{\text{AdS}_2} \int_{\gamma_{12}} \widehat{G}_1\widehat{G}_2G_{12|m}\int_{\gamma_{34}} G_{34|n}\widehat{G}_3\widehat{G}_4 
\\[1.5mm]
\dps
+ \sum_{n}\beta^{34}_n\int_{\text{AdS}_2} \widetilde{G}_1\widehat{G}_2\int_{\gamma_{34}} G_{34|n}\widehat{G}_3\widehat{G}_4 + (1 \leftrightarrow 2) 
\\[1.5mm]
\dps
+ \sum_{n}\beta^{12}_n\int_{\text{AdS}_2} \int_{\gamma_{12}} \widehat{G}_1\widehat{G}_2G_{12|n}\widetilde{G}_3\widehat{G}_4 + (3 \leftrightarrow 4) 
\ea
\vspace{7mm} \\

& \xrightarrow[\hspace{5.7mm}\rm (III)\hspace{5.7mm}]{\eqref{transition_id},\, \eqref{geodesic_split}} & 
\dps 
\ba{l} 
\dps
\hspace{3.5mm}\sum_{n}\rho^{34}_n  \int_{\gamma_{12}} \widehat{G}_1\widehat{G}_2\int_{\gamma_{34}} G_{34|n}\widehat{G}_3\widehat{G}_4 + \sum_{n}\rho^{12}_n \int_{\gamma_{12}} \widehat{G}_1\widehat{G}_2\int_{\gamma_{34}} G_{12|n}\widehat{G}_3\widehat{G}_4 
\\[.5mm]
\dps
+ \sum_{n,m}\beta^{234|n}_m\beta^{34}_n\int_{\gamma_{34}} \cV_{h_{234|nm} h_2 h_{34|n} }\widehat{G}_3\widehat{G}_4 + (1 \leftrightarrow 2) 
\\[.5mm]
\dps
+ \sum_{n,m}\beta^{124|n}_m\beta^{12}_n\int_{\gamma_{12}} \widehat{G}_1\widehat{G}_2\cV_{h_{12|n} h_{124|nm} h_4} + (3 \leftrightarrow 4) 
\ea
\vspace{7mm} \\

& \xrightarrow[\rm (IV)]{\eqref{geodesic_rep},\, \eqref{alt_geodesic_rep}} & 
\dps 
\ba{l} 
\dps
\hspace{3.5mm}\sum_{n}\tilde{\rho}^{34}_n \cV_{h_1h_2h_3h_4, h_{34|n}}+\sum_{n}\tilde{\rho}^{12}_n \cV_{h_1h_2h_3h_4, h_{12|n}} 
\\[.5mm]
\dps
+ \sum_{n,m}\tilde{\beta}^{234|n}_m  \cV_{h_{234|nm}h_2h_3h_4, h_{34|n}}+ (1 \leftrightarrow 2) 
\\[.5mm]
\dps
+ \sum_{n,m}\tilde{\beta}^{124|n}_m  \cV_{h_1h_2h_{124|nm}h_4, h_{12|n}} + (3 \leftrightarrow 4) \,.
\ea
\ea
\ee
where $\beta,\tilde{\beta},\rho, \tilde{\rho}$ are fixed coefficients whose  explicit forms  will be given below; the labels $i$  and $i...j|m...n$ of the (modified) propagators $G_i$ and $G_{i...j|m...n}$  denote the conformal weights  $h_i$ and  $h_i+...+h_j+2m+...+2n$, respectively.  Let us now perform each step of the algorithm \eqref{scheme} consecutively. 

\begin{figure}
\centering
\includegraphics[scale=0.7]{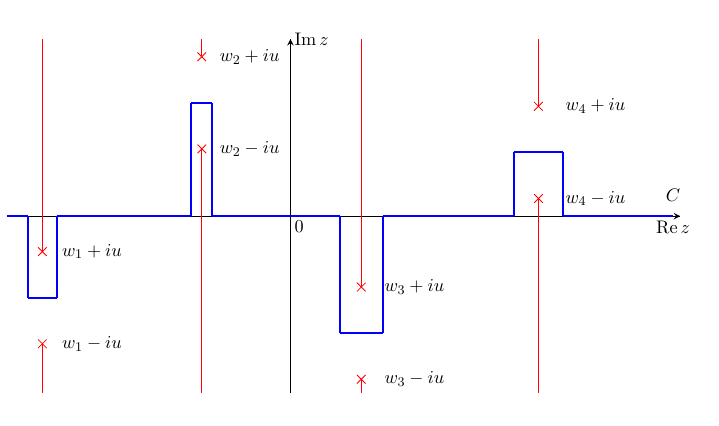}
\caption{Contour $C$ in the complex $z$-plane. Red crosses and lines denote the branch points and  cuts of the integrand in \eqref{cont_mod}.}
\label{fig:C}
\end{figure}

\begin{figure}
\centering
\includegraphics[scale=0.73]{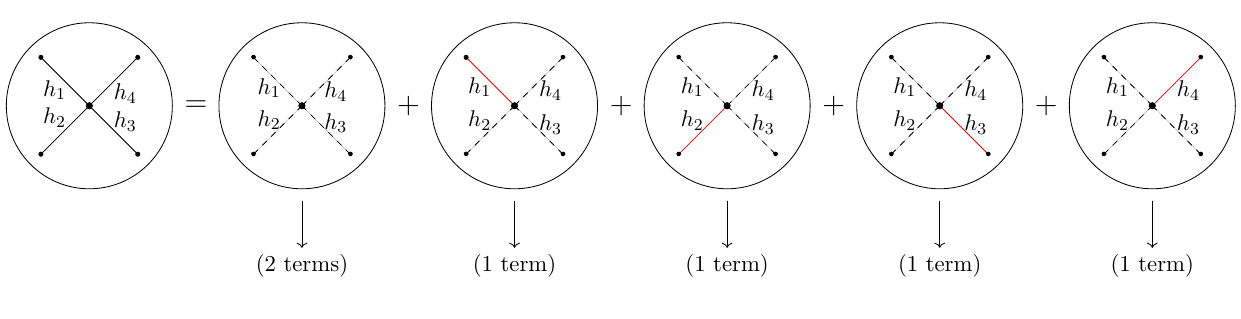}
\caption{Decomposition of a four-point contact \ads Feynman diagram into diagrams with the modified propagators \eqref{G_hat} and  \eqref{G_tilde}.  On the left-hand side, four scalars with masses $m^2_i = h_i(h_i-1)$ are located in the bulk  points $\bx_i$, $i=1,2,3,4$; the central dot denotes the AdS integration, and the black solid lines represent the  standard bulk-to-bulk scalar propagators. On the right-hand side, the red lines denote  the modified propagators $\widetilde{G}_h(\bx,\bx')$, while the dashed lines denote the modified  propagators $\widehat{G}_h(\bx,\bx',w)$. The first diagram in this decomposition  produces two terms in the final expansion \eqref{vert_decomp}, whereas  the other diagrams produce only one. The notation ($N$ terms) indicates the number of terms produced from each corresponding diagram by the remainder of the decomposition algorithm in the final expansion \eqref{vert_decomp}.}
\label{fig:contact_modified}
\end{figure}

{\bf I.} The first step in \eqref{scheme} leads to a five-term decomposition:
\be 
\label{cont_mod}
\ba{l}
\dps
\dc(\bx_i)=\int_{\text{AdS}_2} d^2\bx\;\sqrt{g(\bx)}\; \prod_{i=1}^4 G_{h_i}(\bx,\bx_i) 
\vspace{3mm}
\\
\dps
\hspace{16mm}= \prod_{j=1}^4\int_{z_j-iu_j}^{z_j+iu_j}dw_j\int_0^\infty\frac{du}{u^2}\int_{C}dz\; \prod_{i=1}^4 \widehat{G}_{h_i}(\bx,\bx_i,w_i)+
\vspace{3mm}
\\
\dps
\hspace{16mm}+\pi \sum_{i=1}^4\int_{0}^{u_i} \frac{du}{u^2}\int_{z_i-i(u-u_i)}^{z_i+i(u-u_i)}dz\; \widetilde{G}_{h_i}(\bx,\bx_i) \prod_{\substack{j=1\\j\neq i}}^4 \int_{z_j-iu_j}^{z_j+iu_j}dw_j\;\widehat{G}_{h_j} (\bx,\bx_j,w_j)\,,
\ea
\ee
where the integration contour $C$ is shown in fig.~\bref{fig:C}.  This decomposition, illustrated diagrammatically in fig.~\bref{fig:contact_modified}, is particularly convenient for boundary analysis. Specifically, the boundary asymptotics of the first term on the right-hand side reproduces the four-point contact Witten diagram, while the remaining terms are sub-leading, provided that the triangle inequalities \eqref{triangle_identity1} are satisfied.

\begin{figure}
\centering
\includegraphics[scale=0.9]{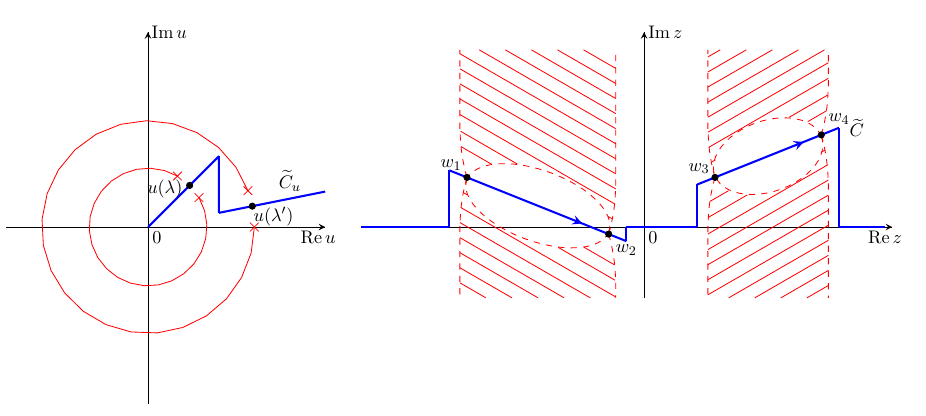}
\caption{Contours $\widetilde{C}_u$ and $\widetilde{C}$ in the complex $u$- and $z$-plane, respectively. Red crosses and lines denote the  branch points and cuts of the integrand in the first term of \eqref{contact_interm_1}. Red shaded areas in the $z$-plane indicate  possible locations of branch points and cuts  for $\lambda,\lambda'\in\RR$.}
\label{fig:tilde_C}
\end{figure}

{\bf II.} Applying the geodesic decomposition identity \eqref{geodesic_prop} to $\widehat{G}_{h_1}\widehat{G}_{h_2}$ and $\widehat{G}_{h_3}\widehat{G}_{h_4}$ yields
$$
\ba{l}
\dps
\dc(\bx_i)=
\prod_{j=1}^4\int_{z_j-iu_j}^{z_j+iu_j}dw_j\int_{\gamma_{12}}d\lambda\;\int_{\gamma_{34}}d\lambda'\; \sum_{n,m=0}^{\infty} \frac{a^{h_1h_2}_n}{\beta_{h_{12|n}h_1h_2}} \frac{a^{h_3h_4}_m}{\beta_{h_{34|m}h_3h_4}} 
\vspace{3mm}
\\
\dps
\times\int_{\widetilde{C}_u}\frac{du}{u^2}\int_{\widetilde{C}}dz\, \widehat{G}_{h_1}(\bx(\lambda),\bx_1,w_1)\widehat{G}_{h_2}(\bx(\lambda),\bx_2,w_2)\widehat{G}_{h_3}(\bx(\lambda'),\bx_3,w_3)\widehat{G}_{h_4}(\bx(\lambda'),\bx_4,w_4)
\vspace{3mm}
\\
\dps
\times G_{h_{12|n}}(\bx(\lambda),\bx) G_{h_{34|m}}(\bx(\lambda'),\bx)
\vspace{3mm}
\\
\dps
+\pi \sum_{m=0}^\infty \frac{a^{h_3h_4}_m}{\beta_{h_{34|m}h_3h_4}} \int_{0}^{u_1} \frac{du}{u^2}\int_{z_1-i(u-u_1)}^{z_1+i(u-u_1)}dz\; \widetilde{G}_{h_1}(\bx,\bx_1)  \int_{z_2-iu_2}^{z_2+iu_2}dw_2\;\widehat{G}_{h_2} (\bx,\bx_2,w_2) 
\vspace{3mm}
\ea
$$
\be
\label{contact_interm_1}
\ba{l}
\dps
\times\prod_{j=3}^4\int_{z_j-iu_j}^{z_j+iu_j}dw_j\int_{\gamma_{34}} d\lambda'\, \widehat{G}_{h_3}(\bx(\lambda'),\bx_3,w_3)\widehat{G}_{h_4}(\bx(\lambda'),\bx_4,w_4) G_{h_{34|m}}(\bx(\lambda'),\bx) + (1\leftrightarrow 2) 
\vspace{3mm}
\\
\dps
+\pi \sum_{n=0}^\infty\frac{a^{h_1h_2}_n}{\beta_{h_{12|n}h_1h_2}}\int_{0}^{u_3} \frac{du}{u^2}\int_{z_3-i(u-u_3)}^{z_3+i(u-u_3)}dz\; \widetilde{G}_{h_3}(\bx,\bx_3)  \int_{z_4-iu_4}^{z_4+iu_4}dw_4\;\widehat{G}_{h_4} (\bx,\bx_4,w_4)
\vspace{3mm}
\\
\dps
\times\prod_{j=1}^2\int_{z_j-iu_j}^{z_j+iu_j}dw_j\int_{\gamma_{12}} d\lambda\,\widehat{G}_{h_1}(\bx(\lambda),\bx_1,w_1)\widehat{G}_{h_2}(\bx(\lambda),\bx_2,w_2) G_{h_{12|n}}(\bx(\lambda),\bx) + (3\leftrightarrow 4)\,,
\ea
\ee
where we have interchanged the integration order in $\lambda$, $\lambda'$, and $\bx$. Additionally, the contours for  $u$ and $z$ integrations have been modified to $\widetilde{C}_u$ and $\widetilde{C}$ (shown in fig.~\bref{fig:tilde_C}) to satisfy the restriction $|\xi(\bx,\bx(\lambda))|\leq 1$ required by the geodesic decomposition identity \eqref{geodesic_prop}. 

{\bf III.} After applying the transition identity \eqref{transition_id} to the first term, we  use  the splitting identity \eqref{geodesic_split} to transform each term containing $\widetilde{G}$ into an infinite sum of three-point AdS vertex functions:
$$
\ba{l}
\dps
\dc(\bx_i)=
\prod_{j=1}^4\int_{z_j-iu_j}^{z_j+iu_j}dw_j\; \sum_{n,m=0}^{\infty} \frac{1}{{\dm^{h_{12|n}}_{h_{34|m}}}}\frac{a^{h_1h_2}_n}{\beta_{h_{12|n}h_1h_2}} \frac{a^{h_3h_4}_m}{\beta_{h_{34|m}h_3h_4}} \int_{\gamma_{12}}d\lambda\;\int_{\gamma_{34}}d\lambda'
\vspace{3mm}
\\
\dps
\times\, \widehat{G}_{h_1}(\bx(\lambda),\bx_1,w_1)\widehat{G}_{h_2}(\bx(\lambda),\bx_2,w_2)G_{h_{12|n}}(\bx(\lambda),\bx(\lambda'))\widehat{G}_{h_3}(\bx(\lambda'),\bx_3,w_3)\widehat{G}_{h_4}(\bx(\lambda'),\bx_4,w_4)
\vspace{3mm}
\\
\dps
+\prod_{j=1}^4\int_{z_j-iu_j}^{z_j+iu_j}dw_j\; \sum_{n,m=0}^{\infty} \frac{1}{{\dm^{h_{34|m}}_{h_{12|n}}}}\frac{a^{h_1h_2}_n}{\beta_{h_{12|n}h_1h_2}} \frac{a^{h_3h_4}_m}{\beta_{h_{34|m}h_3h_4}} \int_{\gamma_{12}}d\lambda\;\int_{\gamma_{34}}d\lambda'\, 
\ea
$$
\be 
\label{contact_decomp}
\ba{l}
\dps
\times\,\widehat{G}_{h_1}(\bx(\lambda),\bx_1,w_1)\widehat{G}_{h_2}(\bx(\lambda),\bx_2,w_2)G_{h_{34|m}}(\bx(\lambda'),\bx(\lambda))\widehat{G}_{h_3}(\bx(\lambda'),\bx_3,w_3)\widehat{G}_{h_4}(\bx(\lambda'),\bx_4,w_4)
\vspace{3mm}
\\
\dps
+\sum_{n,m=0}^{\infty}\frac{a^{h_3h_4}_m}{\beta_{h_{34|m}h_3h_4}}\frac{a^{h_2h_{34|m}}_n}{{\dm^{h_{234|mn}}_{h_1}}}
\prod_{j=3}^4\int_{z_j-iu_j}^{z_j+iu_j}dw_j
\vspace{3mm}
\\
\dps
\times\int_{\gamma_{34}} d\lambda'\,\widehat{G}_{h_3}(\bx(\lambda'),\bx_3,w_3)\widehat{G}_{h_4}(\bx(\lambda'),\bx_4,w_4) \cV_{h_{34|m}h_{234|mn}h_2}(\bx(\lambda'), \bx_1, \bx_2)  + (1\leftrightarrow 2) 
\vspace{3mm}
\\
\dps
+\sum_{n,m=0}^{\infty}\frac{a^{h_1h_2}_n}{\beta_{h_{12|n}h_1h_2}}\frac{a^{h_4h_{12|n}}_m}{{\dm^{h_{124|mn}}_{h_3}}} \prod_{j=1}^2\int_{z_j-iu_j}^{z_j+iu_j}dw_j
\vspace{3mm}
\\
\dps
\times\int_{\gamma_{12}} d\lambda\,\widehat{G}_{h_1}(\bx(\lambda),\bx_1,w_1)\widehat{G}_{h_2}(\bx(\lambda),\bx_2,w_2)\cV_{h_{12|n}h_{124|mn}h_4}(\bx(\lambda), \bx_3, \bx_4) + (3\leftrightarrow 4)\,,
\ea
\ee
where $h_{i_1...i_k|n_1...n_l} = h_{i_1} + ... + h_{i_k} + 2n_1+...+2n_l$.

\begin{figure}
\centering
\includegraphics[scale=0.7]{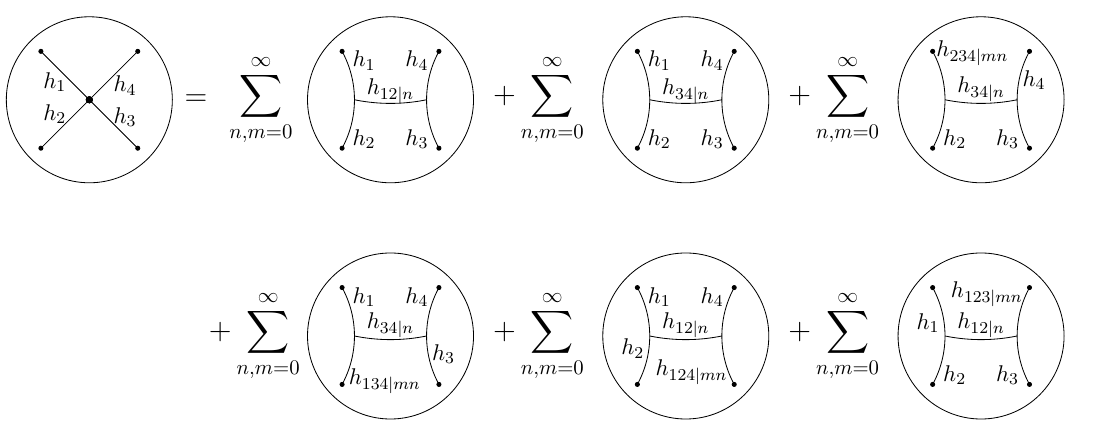}
\caption{Wilson network expansion of a four-point contact \ads Feynman diagram.  On the right-hand side,  the graphs denote matrix elements of Wilson line networks, consisting of  Wilson lines (curved lines) connected by $\sltwo$ intertwiners. The weights are given by $h_{i...j|n...m} = h_i+...+\h_j+2n+...+2m$. }
\label{fig:contact_vertex}
\end{figure}

{\bf IV.} We observe  that the first two terms in \eqref{contact_decomp} take the form of the geodesic integral representation \eqref{geodesic_rep} for  the AdS vertex function with weights $\bm h = (h_1,h_2,h_3,h_4,h)$, where $h$ is either  $h_{12|n}$ or $h_{34|m}$. The triangle inequalities \eqref{triangle_identity1} imply  that $\bm h\in\DD_{12}\cap\DD_{34}$, ensuring  the applicability of representation \eqref{geodesic_rep} to these terms. Furthermore, the remaining four terms match the structure of the integral representation \eqref{alt_geodesic_rep} with $\bm h = (h_i,h_j,h_k,h_{ijk|mn},h_{ij|n})$ for   $(ijk) \in \{ (123),(124),(341),(342)\}$. Since the triangle inequalities \eqref{triangle_identity1} guarantee that $\bm h\in \PP_k$, the  representation \eqref{alt_geodesic_rep} is also valid. Applying \eqref{alt_geodesic_rep} and \eqref{geodesic_rep}  yields the following expansion  of the four-point contact diagram into AdS vertex functions:
$$
\dc(\bx_i)= \sum_{n,m=0}^{\infty} \aleph^{(0)}_{h_1 h_2, h_3 h_4}(n,m)\, 
\cV_{\h_1 \h_2 \h_3 \h_4,\h_{12|n}}(\bx_i)
+\sum_{n,m=0}^{\infty} \aleph^{(0)}_{h_3 h_4, h_1 h_2}(m,n)\,
\cV_{\h_1 \h_2 \h_3 \h_4,\h_{34|n}}(\bx_i)
$$
\be 
\label{vert_decomp}
\ba{l}
\dps
+\sum_{n,m=0}^{\infty}
\aleph^{(1)}_{h_3 h_4, h_2, h_1}(m,n)\,
\cV_{\h_{234|mn} \h_2 \h_3 \h_4,\h_{34|m}}(\bx_i) + (1\leftrightarrow 2) 
\vspace{3mm}
\\
\dps
+\sum_{n,m=0}^{\infty}
\aleph^{(1)}_{h_1 h_2, h_4, h_3}(n,m)\,
\cV_{\h_1 \h_2 \h_{124|mn} \h_4,\h_{12|n}}(\bx_i) + (3\leftrightarrow 4)\,,
\ea
\ee
where we have defined 
\be 
\label{coef}
\ba{l}
\dps
\aleph^{(0)}_{h_i h_j, h_k h_l}(n,m) =  \frac{a^{h_ih_j}_n a^{h_kh_l}_m}{{\dm^{h_{ij|n}}_{h_{kl|m}}}} \frac{\beta_{h_{ij|n} h_k h_l}}{\beta_{h_{kl|m}h_kh_l}}\,,
\vspace{3mm}
\\
\dps
\aleph^{(1)}_{h_i h_j, h_k, h_l}(n,m) =  \frac{a^{h_ih_j}_n a^{h_kh_{ij|n}}_m}{\dm^{h_{ijk|mn}}_{h_{l}}}\,.
\ea
\ee 
Here, the coefficients $a,\beta$, and $\dm$ are given in \eqref{a_beta}. Note that the coefficients of the terms in the first line of \eqref{vert_decomp} can be summed over $m$, since the AdS vertex function therein  is independent of $m$. The reduced coefficients for the first two terms of \eqref{vert_decomp} are given by
\be 
\label{new_aleph}
\ba{l}
\dps
\aleph^{1,\text{reduced}}_{h_1 h_2, h_3 h_4}(n) = \sum_{m=0}^\infty \aleph^{(0)}_{h_1 h_2, h_3 h_4}(n,m)\,,
\vspace{3mm}
\\
\dps
\aleph^{2,\text{reduced}}_{h_1 h_2, h_3 h_4}(n) = \aleph^{1,\text{reduced}}_{h_3h_4, h_1 h_2}(n)\,.
\ea 
\ee 
The sum on the right-hand side is a formally divergent series that arises from an uncontrolled interchange of the integration over $\bx$ and the summation  over $m$ at step II of the decomposition procedure. A standard approach to rendering the series finite is to introduce a regularization factor $(1-\varepsilon)^m$ into the first term of \eqref{contact_interm_1} prior to the interchange. One then rigorously  interchanges the order of integration and summation, repeats steps III and IV and finally  takes the limit $\varepsilon\to 0^+$, see Appendix \bref{app:regularization} for details. Such a regularization cures the divergence of the series  \eqref{new_aleph}:
\be 
\label{c_as_aleph_reg}
\ba{l}
\dps
\aleph^{1,\text{reduced}}_{h_1 h_2, h_3 h_4}(n) := \lim_{\varepsilon\to0^+}\sum_{m=0}^\infty \aleph^{(0)}_{h_1 h_2, h_3 h_4}(n,m)(1-\varepsilon)^m
\vspace{3mm}
\\
\dps
= \frac{(-)^n\pi^\half}{2 n!}\frac{(\h_1)_n(\h_2)_n\Gamma(\frac{h_3+h_4+h_{12|n}}{2})\Gamma(\frac{h_3+h_4-h_{12|n}}{2})\Gamma(\frac{h_3-h_4+h_{12|n}}{2})\Gamma(\frac{-h_3+h_4+h_{12|n}}{2})}{(\h_1+\h_2-\half+n)_n\Gamma(h_3)\Gamma(h_4)\Gamma(h_{12|n})}\,,
\ea
\ee
where the second equality is established via the following identity
\be 
\label{abel_sum}
 \lim_{y\to1^-} \sum_{n=0}^\infty \frac{a^{h_1h_2}_n}{\dm^{h_3}_{h_{12|n}}}\frac{\beta_{h_3h_1h_2}}{\beta_{h_{12|n}h_1h_2}}y^n
= \frac{\pi^\half}{2}\frac{\Gamma(\frac{h_1+h_2+h_3}{2})\Gamma(\frac{h_1+h_2-h_3}{2})\Gamma(\frac{h_1-h_2+h_3}{2})\Gamma(\frac{-h_1+h_2-h_3}{2})}{\Gamma(h_1)\Gamma(h_2)\Gamma(h_3)}\,.
\ee 

Note that the sum \eqref{new_aleph} appears in several works on the conformal block decomposition of Witten diagrams, see e.g. \cite{Hijano:2015zsa,Jepsen:2019svc}.\footnote{An alternative calculation of this sum in the context of Witten diagrams was presented in \cite{Jepsen:2019svc}; their result matches the second line of \eqref{c_as_aleph_reg}. } Notably, the coefficients in the boundary expansion \eqref{4pt_contact_expansion} can be expressed in terms of the reduced coefficients:
\be 
\label{c_as_aleph}
c^{^{(ij|kl)}}_n= \aleph^{1,\text{reduced}}_{h_i h_j, h_k h_l}(n)\,.
\ee 
To maintain consistency with the literature, we write the sum \eqref{new_aleph} in the divergent form in the decomposition \eqref{vert_decomp}, but implicitly assume throughout that this divergent sum is to be understood in terms of the regularized expression \eqref{c_as_aleph_reg}.

The diagrammatic  representation of the Wilson network expansion \eqref{vert_decomp} is shown in fig.~\bref{fig:contact_vertex}. Structurally, the first two terms in the obtained expansion are similar to those from the conformal block expansion of the four-point contact Witten diagram \eqref{4pt_contact_expansion}, while the remaining  terms do not appear in the boundary expansion. In these sub-leading  terms, one of the external weights $h_l$  is replaced  by  the ``multi-trace weight'' $h_{ijk|mn}$.  In the context of Witten diagrams, terms with such weights first appear in the expansion of the five-point Witten diagram \cite{Jepsen:2019svc}.

\begin{figure}
\centering
\includegraphics[scale=0.7]{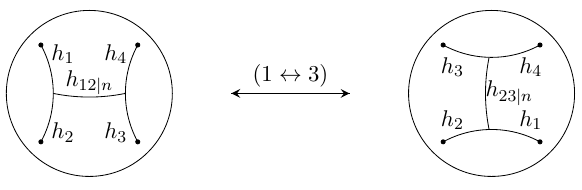}
\caption{Permutation of conformal weights and coordinates in the expansion.  While the left-hand side of \eqref{vert_decomp} is invariant under $(1\leftrightarrow3)$, each individual four-point AdS vertex function on the right-hand side is mapped to an AdS vertex function in the crossed channel.}
\label{fig:crossing}
\end{figure}

Note that the expansion \eqref{vert_decomp}  inherits several permutation symmetries from the contact diagram, which itself is invariant under any permutation of conformal weights and coordinates.   Specifically, \eqref{vert_decomp} is explicitly invariant under the exchanges   $(1\leftrightarrow 2), (3\leftrightarrow 4)$, as well as the simultaneous exchange $(1\leftrightarrow3),  (2\leftrightarrow 4)$. While the four-point contact AdS Feynman diagram is also  symmetric under $(1\leftrightarrow 3)$ or $(2\leftrightarrow 4)$  individually, the right-hand side of \eqref{vert_decomp} is not explicitly so. Such transformations result in switching the channel of the AdS vertex functions (see fig.~\bref{fig:crossing}). The invariance of the Wilson network expansion under these permutations is equivalent to the crossing symmetry.

\subsection{Four-point exchange diagram}

The four-point exchange AdS Feynman diagram \eqref{4pt_diagram_ex_def} can be considered along the same lines. Below we follow the steps of the general decomposition algorithm described in subsection \bref{sec:algorithm}.    

\begin{figure}
\centering
\includegraphics[scale=0.9]{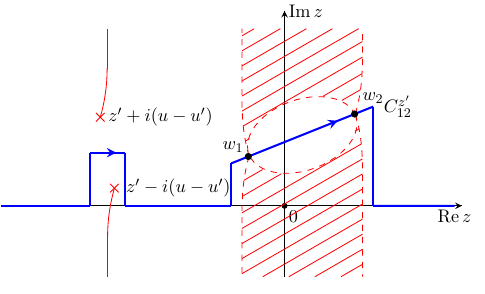}
\caption{Contour $C^{z'}_{12}$ in the complex $z$-plane. Red crosses and lines denote the branch points and cuts of the integrand \eqref{ex_start}. Red shaded areas show possible locations of branch points and cuts of the integrands \eqref{ex_start} and \eqref{exch_int}. The contour $C^{z}_{34}$ in the complex $z'$-plane is defined  by exchanging $\{12\} \leftrightarrow \{34\}$ and $\bx' \leftrightarrow \bx$.}
\label{fig:C_z}
\end{figure}

\begin{figure}
\centering
\includegraphics[scale=0.73]{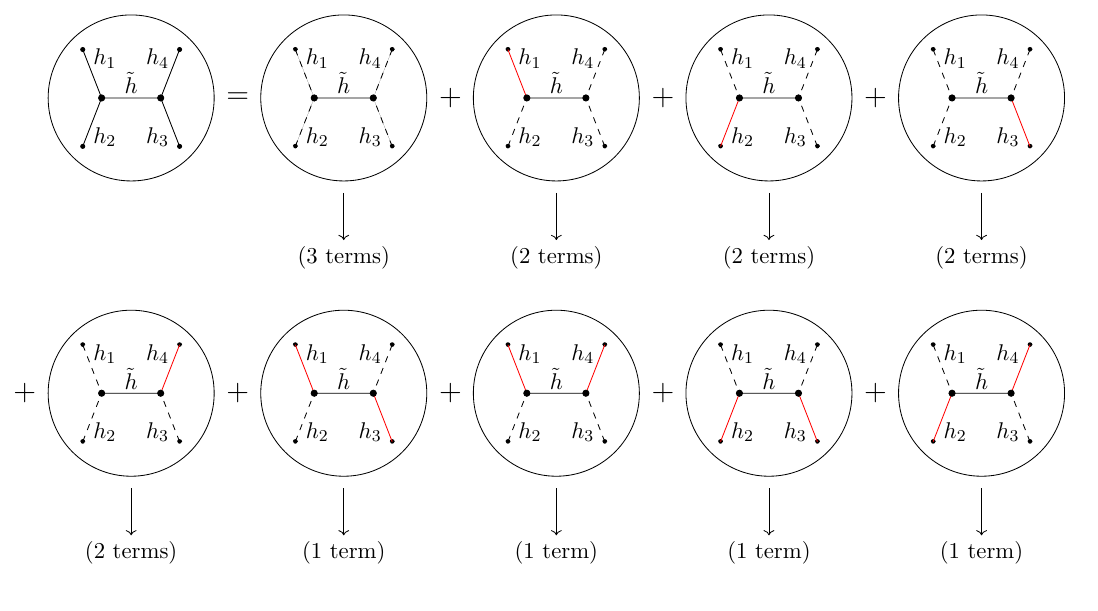}
\caption{Decomposition of a four-point exchange  \ads Feynman  diagram into diagrams with modified propagators \eqref{G_hat} and  \eqref{G_tilde}.  On the left-hand side, four scalars with masses $m^2_i = h_i(h_i-1)$ are located in the bulk  points $\bx_i$, $i=1,2,3,4$, with an intermediate scalar of mass $\widetilde m^2 = \tilde h(\tilde h-1)$ in the exchange channel; the two central dots denote AdS integrations, and the black solid lines represent the standard bulk-to-bulk scalar propagators. On the right-hand side, the red lines denote  modified propagators $\widetilde{G}_h(\bx,\bx')$, while the dashed lines denote  modified  propagators $\widehat{G}_h(\bx,\bx',w)$. The notation ($N$ terms) indicates the number of terms produced from each corresponding diagram by the remainder of the decomposition algorithm in the final expansion \eqref{exch_final}.}
\label{fig:ex_modified}
\end{figure}

{\bf I.} The first step of the algorithm lead to a nine-term decomposition:
\be 
\label{ex_start}
\ba{l}
\dps
\de(\bx_i)=\iint_{\text{AdS}_2} d^2\bx \,d^2\bx'\sqrt{g(\bx)g(\bx')}\; G_{h_1}(\bx,\bx_1)G_{h_2}(\bx,\bx_2)G_{\tilde{h}}(\bx,\bx')G_{h_3}(\bx',\bx_3)  G_{h_4}(\bx',\bx_4) 
\vspace{3mm}
\\
\dps
= 
\prod_{j=1}^4\int_{z_j-iu_j}^{z_j+iu_j}dw_j\int_0^\infty\frac{du}{u^2}\int_{C}dz\int_0^\infty\frac{du'}{u'^2}\int_{C^z_{34}}dz'\;\prod_{i=1}^2 \widehat{G}_{h_i}(\bx,\bx_i,w_i)G_{\tilde{h}}(\bx,\bx')\prod_{k=3}^4 \widehat{G}_{h_k}(\bx',\bx_k,w_k)
\vspace{3mm}
\\
\dps
+\pi \prod_{j=2}^4\int_{z_j-iu_j}^{z_j+iu_j}dw_j
\int_{0}^{u_1} \frac{du}{u^2}\int_{z_1-i(u-u_1)}^{z_1+i(u-u_1)}dz\int_0^\infty\frac{du'}{u'^2}\int_{C^z_{34}}dz'\;
\widetilde{G}_{h_1}(\bx,\bx_1) \widehat{G}_{h_2}(\bx,\bx_2,w_2)G_{\tilde{h}}(\bx,\bx')
\vspace{3mm}
\\
\dps
\times \prod_{i=3}^4 \widehat{G}_{h_i}(\bx',\bx_i,w_i)
+ (1 \leftrightarrow 2) 
+ \pi \prod_{\substack{j=1\\j\neq 3}}^4\int_{z_j-iu_j}^{z_j+iu_j}dw_j\int_{0}^{u_3} \frac{du'}{u'^2}\int_{z_3-i(u'-u_3)}^{z_3+i(u'-u_3)}dz'\int_0^\infty\frac{du}{u^2}\int_{C^{z'}_{12}}dz\;
\\
\dps
\times
\prod_{i=1}^2 \widehat{G}_{h_i}(\bx,\bx_i,w_i)G_{\tilde{h}}(\bx,\bx')\widetilde{G}_{h_3}(\bx',\bx_3) \widehat{G}_{h_4}(\bx,\bx_4,w_4)+ (3 \leftrightarrow 4)
\vspace{3mm}
\\
\dps
+\pi^2\sum_{\substack{i,j=1\\j\neq i}}^2\sum_{\substack{k,l=3\\k\neq l}}^4
\int_{z_j-iu_j}^{z_j+iu_j}dw_j\int_{z_l-iu_l}^{z_l+iu_l}dw_l\int_{0}^{u_i} \frac{du}{u^2}\int_{z_i-i(u-u_i)}^{z_i+i(u-u_i)}dz\int_{0}^{u_k} \frac{du'}{u'^2}\int_{z_k-i(u'-u_k)}^{z_k+i(u'-u_k)}dz'\;
\vspace{3mm}
\\
\dps
\times\,
\widetilde{G}_{h_i}(\bx,\bx_i) \widehat{G}_{h_j}(\bx,\bx_j,w_j) G_{\tilde{h}}(\bx,\bx')\widetilde{G}_{h_k}(\bx',\bx_k) \widehat{G}_{h_l}(\bx,\bx_l,w_l)\,.
\ea
\ee
Here, the contour $C$ is shown in fig.~\bref{fig:C}, while the contours $C^{z'}_{12}$ and $C^{z}_{34}$ are depicted  in fig.~\bref{fig:C_z}. The graphical representation of this decomposition is provided  in fig.~\bref{fig:ex_modified}. Note that the boundary asymptotics of the first term on the right-hand side reproduces the four-point exchange Witten diagram, while the remaining terms are sub-leading, provided that the triangle inequalities \eqref{triangle_identity2} hold.

{\bf II.} Applying the geodesic decomposition identity \eqref{geodesic_prop} to $\widehat{G}_{h_1}\widehat{G}_{h_2}$ and $\widehat{G}_{h_3}\widehat{G}_{h_4}$ in the each term where they appear yields:
$$
\ba{l}
\dps
\de(\bx_i)=
\prod_{j=1}^4\int_{z_j-iu_j}^{z_j+iu_j}dw_j\int_{\gamma_{12}}d\lambda\;\int_{\gamma_{34}}d\lambda'\int_{\widetilde{C}_u}\frac{du}{u^2}\int_{\widetilde{C}}dz\;
\sum_{n,m=0}^{\infty} \frac{a^{h_1h_2}_n}{\beta_{h_{12|n}h_1h_2}} \frac{a^{h_3h_4}_m}{\beta_{h_{34|m}h_3h_4}} 
\ea
$$
\be 
\label{exch_int}
\ba{l}
\dps
\times\int_{\widetilde{C}_u}\frac{du'}{u'^2}\int_{\widetilde{C}^z_{34}}dz'\, \widehat{G}_{h_1}(\bx(\lambda),\bx_1,w_1)\widehat{G}_{h_2}(\bx(\lambda),\bx_2,w_2)\widehat{G}_{h_3}(\bx(\lambda'),\bx_3,w_3)\widehat{G}_{h_4}(\bx(\lambda'),\bx_4,w_4)
\vspace{3mm}
\\
\dps
\times\, G_{h_{12|n}}(\bx(\lambda),\bx) G_{\tilde{h}}(\bx,\bx') G_{h_{34|m}}(\bx(\lambda'),\bx')
\vspace{3mm}
\\
\dps
+\pi \prod_{j=2}^4\int_{z_j-iu_j}^{z_j+iu_j}dw_j\int_{\gamma_{34}} d\lambda'
\int_{0}^{u_1} \frac{du}{u^2}\int_{z_1-i(u-u_1)}^{z_1+i(u-u_1)}dz\;
\widetilde{G}_{h_1}(\bx,\bx_1) \widehat{G}_{h_2}(\bx,\bx_2,w_2)G_{\tilde{h}}(\bx,\bx')
\vspace{3mm}
\\
\dps
\times \sum_{m=0}^\infty \frac{a^{h_3h_4}_m}{\beta_{h_{34|m}h_3h_4}} \int_{\widetilde{C}_u}\frac{du'}{u'^2}\int_{C^z_{34}}dz'\, \widehat{G}_{h_3}(\bx(\lambda'),\bx_3,w_3)\widehat{G}_{h_4}(\bx(\lambda'),\bx_4,w_4) G_{h_{34|m}}(\bx(\lambda'),\bx')
+ (1 \leftrightarrow 2) 
\vspace{3mm}
\\
\dps
+ \pi \prod_{\substack{j=1\\j\neq 3}}^4\int_{z_j-iu_j}^{z_j+iu_j}dw_j\int_{0}^{u_3} \frac{du'}{u'^2}\int_{z_3-i(u'-u_3)}^{z_3+i(u'-u_3)}dz'\int_{\gamma_{12}} d\lambda\; G_{\tilde{h}}(\bx,\bx')\widetilde{G}_{h_3}(\bx',\bx_3) \widehat{G}_{h_4}(\bx,\bx_4,w_4)
\vspace{3mm}
\\
\dps
\times
 \sum_{n=0}^\infty\frac{a^{h_1h_2}_n}{\beta_{h_{12|n}h_1h_2}}\int_{\widetilde{C}_u}\frac{du}{u^2}\int_{C^{z'}_{12}}dz\,\widehat{G}_{h_1}(\bx(\lambda),\bx_1,w_1)\widehat{G}_{h_2}(\bx(\lambda),\bx_2,w_2) G_{h_{12|n}}(\bx(\lambda),\bx)+ (3 \leftrightarrow 4)
\vspace{3mm}
\\
\dps
+\pi^2\sum_{\substack{i,j=1\\j\neq i}}^2\sum_{\substack{k,l=3\\k\neq l}}^4
\int_{z_j-iu_j}^{z_j+iu_j}dw_j\int_{z_l-iu_l}^{z_l+iu_l}dw_l\int_{0}^{u_i} \frac{du}{u^2}\int_{z_i-i(u-u_i)}^{z_i+i(u-u_i)}dz\int_{0}^{u_k} \frac{du'}{u'^2}\int_{z_k-i(u'-u_k)}^{z_k+i(u'-u_k)}dz'\;
\vspace{3mm}
\\
\dps
\times\,
\widetilde{G}_{h_i}(\bx,\bx_i) \widehat{G}_{h_j}(\bx,\bx_j,w_j) G_{\tilde{h}}(\bx,\bx')\widetilde{G}_{h_k}(\bx',\bx_k) \widehat{G}_{h_l}(\bx,\bx_l,w_l)\,.
\ea
\ee

\begin{figure}
\centering
\includegraphics[scale=0.6]{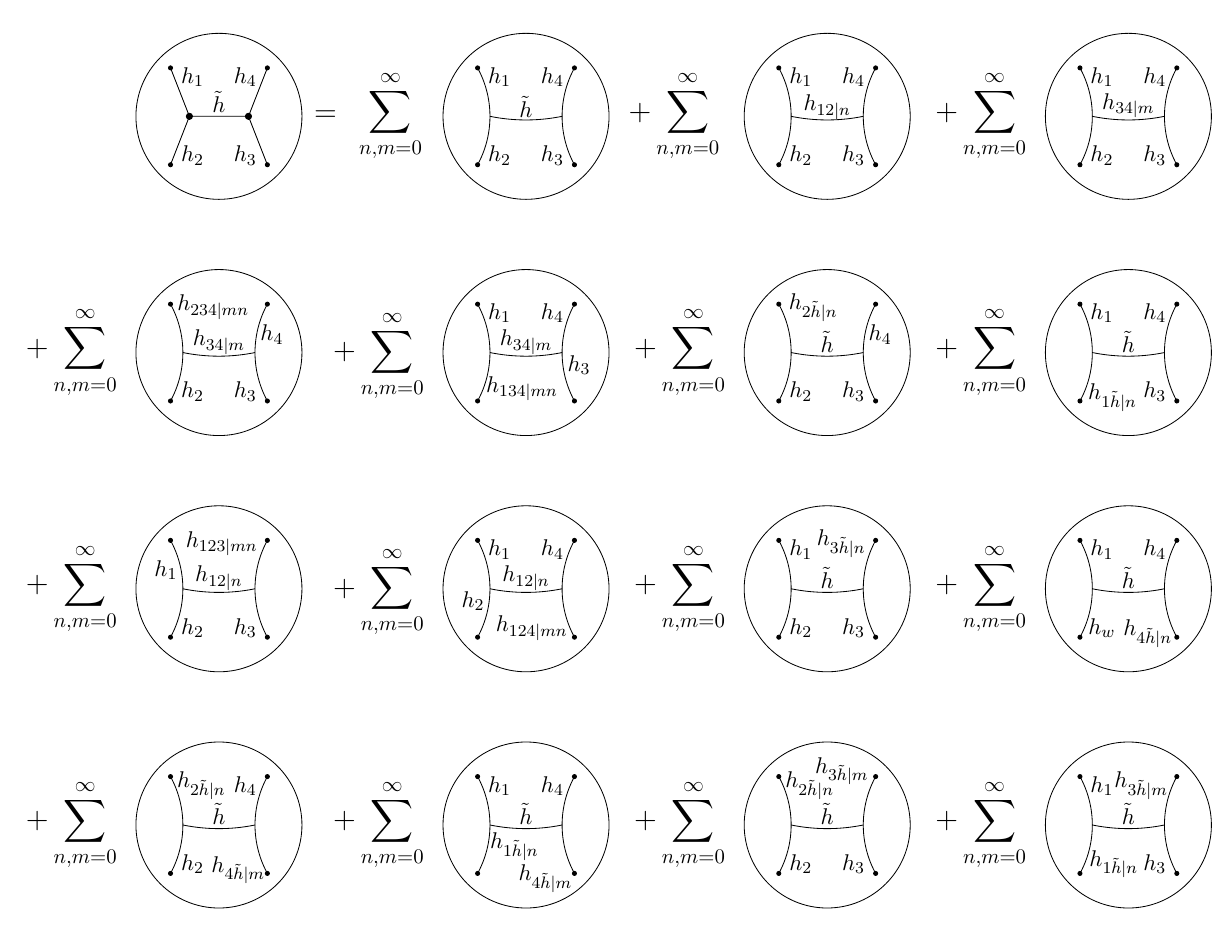}
\caption{Wilson line network decomposition of a four-point exchange \ads Feynman  diagram.  On the right-hand side,  the graphs denote matrix elements of Wilson line networks, consisting of  Wilson lines (curved lines) connected by $\sltwo$ intertwiners. The conformal weights are given by $h_{i...j|n...m} = h_i+...+\h_j+2n+...+2m$. Each line in the right-hand side is symmetric under permutation of conformal weights and coordinates:  $(1\leftrightarrow 2)$, $(3\leftrightarrow 4)$. }
\label{fig:ex_vertex}
\end{figure}

{\bf III-IV.} The rest of the decomposition procedure involves three operations: (1) apply the transition identity \eqref{transition_id} twice to the first term of \eqref{exch_int} and once to each of the subsequent four terms; (2) use the splitting identity \eqref{geodesic_split} to replace integrals of the form $\int G \widehat{G} \widetilde{G}$ with those  of the form $\int G \widehat{G} \widehat{G}$ in every  term containing $\widetilde{G}$; (3) replace each term in the resulting sum with the four-point AdS vertex functions using their integral representations \eqref{alt_geodesic_rep}, \eqref{non_triangle_rep}, or \eqref{geodesic_rep}. The final result is: 
$$
\ba{rcl}
\dps \de(\bx_i) & = & \dps
\overset{{}_{(1)}}{\sum_{n,m=0}^{\infty} \beth^{(0)}_{h_1 h_2 h_3 h_4| \tilde{h}, h_{12|n} h_{34|m}}(n,m)
\cV_{h_1 h_2 h_3 h_4,\tilde{h}}(\bx_1,\bx_2,\bx_3,\bx_4)}
\vspace{3mm}
\\
& + &  \overset{{}_{(2)}}{\dps \sum_{n,m=0}^{\infty} \beth^{(0)}_{h_1 h_2 h_3 h_4| h_{12|n}, \tilde{h} h_{34|m}}(n,m)
\cV_{h_1 h_2 h_3 h_4,h_{12|n}}(\bx_1,\bx_2,\bx_3,\bx_4)}
\vspace{3mm}
\\
\dps
& + &  \overset{{}_{(3)}}{\dps \sum_{n,m=0}^{\infty} \beth^{(0)}_{h_1 h_2 h_3 h_4| h_{34|m}, h_{12|n} \tilde{h}}(n,m)
\cV_{h_1 h_2 h_3 h_4,h_{34|m}}(\bx_1,\bx_2,\bx_3,\bx_4)}
\vspace{3mm}
\\
& + &  \dps \overset{{}_{(4)}}{\sum_{n,m=0}^\infty\beth^{(1)}_{h_3 h_4, h_2, h_1| h_{34|m}, \tilde{h}}(m,n)
\cV_{h_{234|mn} h_2 h_3 h_4,h_{34|m}}(\bx_1,\bx_2,\bx_3,\bx_4)}+  \overset{{}_{(5)}}{\vphantom{\sum_{n,m=0}^\infty}(1 \leftrightarrow 2)}
\ea
$$
\be
\label{exch_final}
\ba{rcl}
\dps
& + & \dps \overset{{}_{(6)}}{\sum_{n,m=0}^\infty\beth^{(1)}_{h_3 h_4, h_2, h_1| \tilde{h}, h_{34|m}}(m,n)
\cV_{h_{2\tilde{h}|n} h_2 h_3 h_4,\tilde{h}}(\bx_1,\bx_2,\bx_3,\bx_4)}
+  \overset{{}_{(7)}}{\vphantom{\sum_{n,m=0}^\infty}(1 \leftrightarrow 2)}
\vspace{3mm}
\\
& + & \dps  \overset{{}_{(8)}}{\sum_{n,m=0}^\infty\beth^{(1)}_{h_1 h_2, h_4, h_3|h_{12|n}, \tilde{h}}(n,m)
\cV_{h_1 h_2 h_{124|mn} h_4,h_{12|n}}(\bx_1,\bx_2,\bx_3,\bx_4)}+  \overset{{}_{(9)}}{\vphantom{\sum_{n,m=0}^\infty}(3 \leftrightarrow 4)}
\vspace{3mm}
\\
\dps
& + & \dps  \overset{{}_{(10)}}{\sum_{n,m=0}^\infty
\beth^{(1)}_{h_1 h_2, h_4, h_3|\tilde{h}, h_{12|n}}(n,m)
\cV_{h_1 h_2 h_{4\tilde{h}|m} h_4,\tilde{h}}(\bx_1,\bx_2,\bx_3,\bx_4)}
+ \overset{{}_{(11)}}{\vphantom{\sum_{n,m=0}^\infty}(3 \leftrightarrow 4)}
\vspace{3mm}
\\
\dps
& + & \dps \overset{{}_{(12-15)}}{\sum_{\substack{i,j=1\\j\neq i}}^2\sum_{\substack{k,l=3\\k\neq l}}^4
\sum_{n,m=0}^\infty\beth^{(2)}_{h_i, h_j| h_k, h_l| \tilde{h}}(n,m)
\cV_{h_{i\tilde{h}|n} \h_i h_{k\tilde{h}|m} \h_k,\tilde{h}}(\bx_j,\bx_i,\bx_l,\bx_k)}\,,
\ea
\ee
where we numbered the terms using the superscript $(i)$ and  defined
\be 
\label{coef_ex}
\ba{l}
\dps
\beth^{(0)}_{h_i h_j h_k h_l| h_s, h_p h_q}(n,m) = \beta_{h_s h_i h_j}\beta_{h_sh_k h_l}\frac{a^{h_i h_j}_n}{\dm^{h_s}_{h_p}\beta_{h_{ij|n}h_i h_j}} \frac{a^{h_k h_l}_m}{\dm^{h_s}_{h_q}\beta_{h_{kl|m}h_k h_l}}\,,
\vspace{3mm}
\\
\dps
\beth^{(1)}_{h_i h_j, h_k, h_l| h_s, h_p}(n,m) = \beta_{h_s h_i h_j}\frac{a^{h_i h_j}_n}{\dm^{h_s}_{h_p}\beta_{h_{ij|n}h_i h_j}} \frac{a^{h_k h_s}_m}{\dm^{h_{ks|m}}_{h_l}}\,,
\vspace{3mm}
\\
\dps
\beth^{(2)}_{h_i, h_j| h_k, h_l| h_s}(n,m) = \frac{a^{h_i h_s}_n}{\dm^{h_{is|n}}_{h_j}} \frac{a^{h_k h_s}_m}{\dm^{h_{ks|m}}_{h_l}}\,.
\ea
\ee 
Here, the coefficients $a,\beta$, and $\dm$ are given in \eqref{a_beta}. The graphical representation of this expansion is shown in fig.~\bref{fig:ex_vertex}. Similar to the case of the contact diagram, some coefficients in the resulting expansion \eqref{exch_final} can be summed over $n$ and $m$, or both:
$$
\ba{l}
\dps
\beth^{1, \text{reduced}} := \lim_{\varepsilon_1\to0^+}\lim_{\varepsilon_2\to0^+}\sum_{n,m=0}^\infty (1-\varepsilon_1)^n(1-\varepsilon_2)^m\,\beth^{(0)}_{h_1 h_2 h_3 h_4| \tilde{h}, h_{12|n} h_{34|m}}(n,m)
\vspace{4mm}
\\
\dps
= \frac{\pi}{4}\frac{\Gamma(\frac{h_1+h_2+\tilde{h}}{2})\Gamma(\frac{h_1+h_2-\tilde{h}}{2})\Gamma(\frac{h_1-h_2+\tilde{h}}{2})\Gamma(\frac{-h_1+h_2+\tilde{h}}{2})}{\Gamma(h_1)\Gamma(h_2)\Gamma(\tilde{h})}
\frac{\Gamma(\frac{h_3+h_4+\tilde{h}}{2})\Gamma(\frac{h_3+h_4-\tilde{h}}{2})\Gamma(\frac{h_3-h_4+\tilde{h}}{2})\Gamma(\frac{-h_3+h_4+\tilde{h}}{2})}{\Gamma(h_3)\Gamma(h_4)\Gamma(\tilde{h})}\,,
\vspace{4mm}
\\
\dps
\beth^{2, \text{reduced}}_{h_1h_2,h_3h_4}(n) := \lim_{\varepsilon\to0^+}\sum_{m=0}^\infty  \, (1-\varepsilon)^m\,\beth^{(0)}_{h_1 h_2 h_3 h_4| h_{12|n}, \tilde{h} h_{34|m}}(n,m)
\vspace{4mm}
\\
\dps
=\frac{(-)^n\pi}{n!}\frac{\Gamma(\tilde{h}+\half)(\h_1)_n(\h_2)_n\Gamma(\frac{h_3+h_4+h_{12|n}}{2})\Gamma(\frac{h_3+h_4-h_{12|n}}{2})\Gamma(\frac{h_3-h_4+h_{12|n}}{2})\Gamma(\frac{-h_3+h_4+h_{12|n}}{2})}{\Gamma(\tilde{h})(h_{12|n}(h_{12|n}-1) - \tilde{h}(\tilde{h}-1))(\h_1+\h_2-\half+n)_n\Gamma(h_3)\Gamma(h_4)\Gamma(h_{12|n})} \,,
\ea
$$
\be 
\label{new_beth}
\ba{l}
\dps
\beth^{3, \text{reduced}}_{h_1h_2,h_3h_4}(m) := \lim_{\varepsilon\to0^+}\sum_{n=0}^\infty  \, (1-\varepsilon)^n\,\beth^{(0)}_{h_1 h_2 h_3 h_4| h_{34|m}, h_{12|n} \tilde{h}}(n,m) = \beth^{2, \text{reduced}}_{h_3h_4,h_1h_2}(m)\,, 
\vspace{4mm}
\\
\dps
\beth^{6, \text{reduced}}_{h_1h_2,h_3h_4}(n) := \lim_{\varepsilon\to0^+}\sum_{m=0}^\infty  \, (1-\varepsilon)^m\,\beth^{(1)}_{h_3 h_4, h_2, h_1| \tilde{h}, h_{34|m}}(m,n)
\vspace{4mm}
\\
\dps
\hspace{24mm}= \frac{\Gamma(h_1+\half)\Gamma(\tilde{h})(h_{12|n}(h_{12|n}-1) - \tilde{h}(\tilde{h}-1))}{\Gamma(\tilde{h}+\half)\Gamma(h_1)(h_{2\tilde{h}|n}(h_{2\tilde{h}|n}-1) - \tilde{h}(\tilde{h}-1))}\,\beth^{2, \text{reduced}}_{h_1h_2,h_3h_4}(n)\,,
\vspace{4mm}
\\
\dps
\beth^{7, \text{reduced}}_{h_1h_2,h_3h_4}(n) := \lim_{\varepsilon\to0^+}\sum_{m=0}^\infty  \, (1-\varepsilon)^m\,\beth^{(1)}_{h_3 h_4, h_1, h_2| \tilde{h}, h_{34|m}}(m,n) = \beth^{6, \text{reduced}}_{h_2h_1,h_3h_4}(n)\,,
\vspace{4mm}
\\
\dps
\beth^{10, \text{reduced}}_{h_1h_2,h_3h_4}(m) :=  \lim_{\varepsilon\to0^+}\sum_{n=0}^\infty  \, (1-\varepsilon)^n\,\beth^{(1)}_{h_1 h_2, h_4, h_3|\tilde{h}, h_{12|n}}(n,m) = \beth^{6, \text{reduced}}_{h_3h_4,h_1h_2}(m)\,, 
\vspace{4mm}
\\
\dps
\beth^{11, \text{reduced}}_{h_1h_2,h_3h_4}(m) :=  \lim_{\varepsilon\to0^+}\sum_{n=0}^\infty  \, (1-\varepsilon)^n\,\beth^{(1)}_{h_1 h_2, h_3, h_4|\tilde{h}, h_{12|n}}(n,m) =\beth^{10, \text{reduced}}_{h_1h_2,h_4h_3}(m)\,, 
\ea 
\ee 
where the coefficient $\beth^{i,\text{reduced}}$ corresponds to the $i$-th term of the expansion  and we used \eqref{abel_sum} to evaluate the coefficients.  We have regularized the arising divergent series using the same procedure as discussed below \eqref{new_aleph}, see Appendix \bref{app:regularization} for details. Note that the group of terms $(12-15)$ in \eqref{exch_final} needs not to be regularized.

The reduced coefficients $\beth^{i,\text{reduced}}$ here play the same role as the reduced coefficient $\aleph^{1,\text{reduced}}$ in the expansion of the contact diagram \eqref{vert_decomp}, i.e. the coefficients in the boundary expansion \eqref{4pt_exchange_expansion} can be constructed from the reduced coefficients $\beth^{i,\text{reduced}}$ (cf. \eqref{c_as_aleph}):
\be 
\label{c_as_beth}
\ba{l}
\dps
c_0 = \beth^{1, \text{reduced}}\,,
\vspace{3mm}
\\
\dps
c^{^{(ij|kl)}}_n= \dm^{h_{ij|n}}_{\tilde{h}}\beth^{2, \text{reduced}}_{h_ih_j,h_kh_l}(n) \,.
\ea
\ee 

Similar to the contact AdS Feynman diagram, the resulting expansion \eqref{exch_final} possesses  permutation symmetries inherited from the original exchange AdS Feynman diagram. Specifically, the right-hand side of \eqref{exch_final} is invariant under the exchanges of weights and coordinates  $(1\leftrightarrow 2)$ and  $(3\leftrightarrow 4)$ as well as the simultaneous exchange of $(1\leftrightarrow3)$ and  $ (2\leftrightarrow 4)$. Note that the individual exchanges $(1\leftrightarrow3)$ or $(2\leftrightarrow4)$ are symmetries of neither the original diagram nor the expansion itself.

\subsection{Conformal boundary asymptotics}

Let us show that the leading boundary asymptotics of the Wilson network expansions for four-point AdS Feynman diagrams \eqref{vert_decomp} and \eqref{exch_final} recover the conformal block expansions \eqref{4pt_contact_expansion} and \eqref{4pt_exchange_expansion}: 
\be
\ba{c}
\dc(\bx_i) \simeq u^{\sum_i h_i}\dc(z_i)\quad \text{as}\;\; u \to 0\,,
\vspace{3mm}
\\
\de(\bx_i) \simeq u^{\sum_i h_i} \de(z_i)\quad \text{as}\;\; u \to 0\,,
\ea
\ee  
where $u\equiv u_1 = ... = u_4$ is a common radial coordinate approaching  the conformal boundary. These asymptotic expansions are guaranteed by the triangle identities \eqref{triangle_identity1} and \eqref{triangle_identity2}. The recovery of the conformal blocks is achieved by employing the extrapolate dictionary relation for the four-point AdS vertex function \cite{Alkalaev:2023axo}:
\be
\label{extrapol}
\lim_{u\to 0}u^{-\sum_{i=1}^4\h_i}\,\cV_{h,\tilde{h}}(\bx_1,...,\bx_4)\Big|_{u_1=...=u_4}= \,F_{h, \tilde{h}}(z_1,...,z_4)\,,
\ee
where $F_{h,h'}(z_1,...,z_4)$ is the four-point scalar conformal block. Consequently, each AdS vertex function in the Wilson network expansions can be systematically replaced by its corresponding conformal block in the boundary limit.

\paragraph{Contact diagram.} Using \eqref{extrapol} and \eqref{c_as_aleph}, we obtain  the boundary asymptotics as $u\to0$ of the Wilson network  expansion \eqref{vert_decomp}: 
$$
\ba{l}
\dps
u^{h}\dc(z_i)\simeq  u^{h}\bigg\{\sum_{n=0}^{\infty}c^{^{(12|34)}}_n \,F_{\h_1 \h_2 \h_3 \h_4, \h_{12|n}}(z_1,...,z_4)
+\sum_{n=0}^{\infty}c^{^{(34|12)}}_n\,F_{\h_1 \h_2 \h_3 \h_4, \h_{34|n}}(z_1,...,z_4)\bigg\}
\ea
$$
\be 
\label{vert_decomp_bndry}
\ba{l}
\dps
+u^{2(h_2+h_3+h_4)}\,
 \frac{1}{\dm^{h_{234}}_{h_1}}
\,F_{\h_{234} \h_2 \h_3 \h_4, \h_{34}}(z_1,...,z_4) + (1\leftrightarrow 2) 
\vspace{3mm}
\\
\dps
+u^{2(h_1+h_2+h_4)}\,
\frac{1}{\dm^{h_{124}}_{h_3}}
\,
F_{\h_1 \h_2 \h_{124} \h_4, \h_{12}}(z_1,...,z_4) + (3\leftrightarrow 4)\,,
\ea
\ee
 where we have kept only the leading asymptotics for  each term  and introduced the notation $h_{ijk} = h_i+h_j+h_k$ and $ h = h_1+h_2+h_3+h_4$. Note that the first two terms here are dominant  only if the conformal weights satisfy  the  triangle inequalities \eqref{triangle_identity1}. For instance, let us consider the third term and subtract its power of $u$ from that  of the first term:
\be 
h -2(h_2+h_3+h_4) = h_1-h_2-h_3-h_4 < 0 \;\;\Rightarrow\;\; h < 2(h_2+h_3+h_4)\,,
\ee 
where we have used  the triangle inequality $-h_1+h_2+h_3+h_4 > 0$ from \eqref{triangle_identity1}. Whenever this condition holds,   \eqref{vert_decomp_bndry} coincides with the conformal block expansion  \eqref{4pt_contact_expansion}. Otherwise, the boundary asymptotics contains additional terms with conformal weights  given by  partial sums of original external weights.

\paragraph{Exchange diagram.} Note that the sums of the external weights, which fix the $u$-dependence of the AdS vertex function as $u\to0$ \eqref{extrapol},  are identical  and minimal for the first three terms of the expansion \eqref{exch_final} compared to the others, provided that the triangle inequalities \eqref{triangle_identity2} hold. Consequently, these  first three terms are leading as $u\to 0$. Omitting the sub-leading terms and using \eqref{c_as_beth} to rewrite the coefficients, we obtain:  
\be 
\label{vert_ex_bndry}
\ba{l}
\dps
u^{h}\de(z_i)\simeq  u^{h}c_0\, F_{h,\tilde{h}}(z_1,...,z_4)
\vspace{3mm}
\\
\dps
\hspace{11mm}+u^{h}\bigg\{\sum_{n=0}^{\infty}\frac{c^{^{(12|34)}}_n}{\dm^{h_{12|n}}_{\tilde{h}}}\, F_{h,h_{12|n}}(z_1,...,z_4)+ \sum_{n=0}^{\infty}\frac{c^{^{(34|12)}}_n}{\dm^{h_{34|n}}_{\tilde{h}}}\, F_{h,h_{34|n}}(z_1,...,z_4)\bigg\}\,,
\ea
\ee
which coincides with the expansion of the four-point exchange Witten diagram \eqref{4pt_exchange_expansion}.\footnote{Note that the triangle inequalities in both cases, \eqref{triangle_identity1} and \eqref{triangle_identity2}, are necessary not only for  consistency of the results but also for the convergence of the Witten diagrams. When these triangle inequalities are not satisfied, regularization of the integrals becomes necessary; see \cite{Castro:2024cmf} for details regarding the three-point Witten diagram.}

\section{Conclusion}
\label{sec:conclusion}

In this paper, we have shown that the Wilson line network expansion can be explicitly constructed for both four-point contact and exchange AdS$_2$ Feynman diagrams. We have demonstrated that their boundary asymptotics  recover the well-established conformal block decompositions of Witten diagrams. In this regard, the three-point case considered earlier is  is not sufficiently representative, as all three-point AdS vertex functions reduce to fixed power-law functions of the boundary coordinates. In contrast, the four-point case provides a much more stringent test, as it involves a non-trivial dependence on the conformal cross-ratios. 

A notable feature of this construction is that individual contributions  appearing in the Wilson network expansion  may involve  representations  at the endpoints corresponding to multi-trace primary operators; however, these contributions are suppressed near the conformal boundary, ensuring agreement with the expected boundary behavior.

Looking ahead, the extension of this formalism to $n$-point diagrams will likely require additional integral identities as well as AdS vertex functions in channels beyond the comb channel. Ultimately, it is expected that a closed system of identities exists, which would allow any AdS Feynman diagram to be expanded into Wilson networks.

\vspace{3mm} 

\noindent \textbf{Acknowledgements.}  We are  grateful to Wladyslaw Wachowski for useful discussions. Our work was supported by the Foundation for the Advancement of Theoretical Physics and Mathematics “BASIS”.

\appendix

\section{Derivation  of propagator identities}
\label{app:derivation}

\subsection{Geodesic decomposition identity \eqref{geodesic_prop}}
\label{app:geod_dec}

To prove the geodesic decomposition identity, we first consider its boundary version \cite{Hijano:2015zsa}:
\be 
\label{bndry_geodesic_prop}
K_{h_1}(\bx,z_1)K_{h_2}(\bx,z_2) 
= \sum_{n=0}^\infty \frac{a^{h_1h_2}_n}{\beta_{h_{12|n}h_1h_2}} 
\int_{\gamma_{12}}d\lambda\, K_{h_1}(\bx(\lambda),z_1) K_{h_2}(\bx(\lambda),z_2) G_{h_{12|n}}(\bx(\lambda),\bx)\,,
\ee
where $\gamma_{12}$ is the geodesic $\bx = \bx(\lambda)$ connecting two boundary points $z_1, z_2\in\RR$ \eqref{comp_geod} 
\be 
\label{real_geod}
u(\lambda) = \frac{|z_1-z_2|}{2\cosh\lambda}\,,
\qquad
z(\lambda) = \frac{(z_1+z_2)}{2} + \frac{(z_1-z_2)}{2}\tanh\lambda\,.
\ee 

The bulk version \eqref{geodesic_prop} of this identity can be obtained by multiplying both sides of \eqref{bndry_geodesic_prop} by the smearing functions $\mathbb{K}_{h_1}(\bx_1,z_1)$ and $\mathbb{K}_{h_2}(\bx_2,z_2)$ \eqref{smear} and integrating over the complex boundary points $z_1, z_2\in\CC$. To this end, we first need to extend the relation   \eqref{bndry_geodesic_prop} to the complex plane, $z_1, z_2\in\CC$. The only difficulty here lies in  defining the geodesic $\gamma_{12}$ \eqref{real_geod} for $z_1, z_2\in\CC$, as it involves the absolute value $|z_1-z_2|$ in $u(\lambda)$, which does not have a unique analytic extension. This is a crucial point,  since the propagators evaluated along the geodesics depend on powers of $u(\lambda)$.  We verify by explicit calculation that the extension of \eqref{bndry_geodesic_prop} to $z_1, z_2\in\CC$ is obtained by replacing  the geodesic  \eqref{real_geod} with \eqref{comp_geod}.  

By substituting the propagators  $K_{h_i}(\bx(\lambda),z_i)$ \eqref{bulk-to-boundary} and  $G_{h_{12|n}}(\bx(\lambda),\bx)$ \eqref{bulk-to-bulk}, the geodesic $\gamma_{12}$ \eqref{comp_geod},  and the coefficients $a^{h_1h_2}_n, \beta_{h_{12|n}h_1h_2}$ \eqref{a_beta} into \eqref{bndry_geodesic_prop}, and assuming that $u,z,z_1,z_2\in\CC$, we obtain:
\be
\label{interm_geod}
\ba{l}
\dps
K_{h_1}(\bx,z_1)K_{h_2}(\bx,z_2) 
= \frac{2(z_1-z_2)^{-h_1-h_2}}{\Gamma(h_1)\Gamma(h_2)}\sum_{n=0}^\infty \frac{(-)^n}{n!(\h_1+\h_2-\half+n)_n} 
\vspace{3mm}
\\
\dps
\times 
\int_{\RR}d\lambda\, e^{\lambda (h_1-h_2)}\sum_{k=0}^\infty\frac{\Gamma(h_{12|n}+2k)}{k!(h_{12|n}+\half)_k}2^{-h_{12|n}-2k}\xi(\bx,\bx(\lambda))^{h_{12|n}+2k}
\,,
\ea
\ee
where $\xi(\bx,\bx(\lambda))$ is defined in \eqref{bulk-to-bulk} and the Gauss hypergeometric function in the bulk-to-bulk propagator has been expanded into a series. The sum over $k$ converges as long as  $|\xi(\bx,\bx(\lambda))| \leq 1$ for all $\lambda\in\RR$. By  redefining  the summation indices  $n=m-k$ and performing the summation over $k$, we obtain 
\be
\ba{l}
\dps
K_{h_1}(\bx,z_1)K_{h_2}(\bx,z_2) 
= \frac{(z_1-z_2)^{-h_1-h_2}}{\Gamma(h_1)\Gamma(h_2)}\sum_{m=0}^\infty \frac{(-)^m2^{1-h_{12|m}}\Gamma(h_{12|m})}{m!(\h_1+\h_2-\half+m)_m} 
\vspace{3mm}
\\
\dps
\times
\int_{\RR}d\lambda\, \frac{\xi(\bx,\bx(\lambda))^{h_{12|m}}}{e^{\lambda (h_2-h_1)}}\,
{}_3F_2\left[\half-\h_{12|m}, -m, \frac{5}{4}-\frac{\h_{12|m}}{2}; \frac{1}{4}-\frac{\h_{12|m}}{2}, \frac{3}{2}-\h_{12|m}+m\Big|\,1\right].
\ea
\ee
The hypergeometric function ${}_3F_2$ vanishes  for any $m>0$ due to the following identity (a special case of Dixon's theorem) \cite{Bateman:100233}:
\be 
{}_3F_2\left[a, b,  \frac{a}{2}+1; \frac{a}{2}, a-b+1\Big|\,1\right] = 0, \quad a \neq 0 \wedge \re(b) < 0\,,
\ee 
as in our case $a = \half-\h_{12|m}\neq 0$ due to the restriction $h_i\geq\half$. Therefore, the only non-zero term in the sum \eqref{interm_geod} over $m$ is the first one $m=0$. Changing the variable to $e^{2\lambda} = \frac{t}{1-t}$ and integrating over $t$ yields:
\be
\ba{l}
\dps
K_{h_1}(\bx,z_1)K_{h_2}(\bx,z_2) 
= K_{h_1+h_2}(\bx,z_2)\,{}_2F_1\left[h_1+h_2, h_1; h_1+h_2\Big|\,\frac{(z_2-z)^2-(z_1-z)^2}{u^2+(z-z_2)^2}\right].
\ea
\ee
Using  the property  ${}_2F_1\left[a, b; a\big|\,x\right] = (1-x)^{-b}$ 
completes the proof of the identity \eqref{bndry_geodesic_prop} for  complex points $u,z,z_1,z_2\in\CC$ provided  that  $|\xi(\bx,\bx(\lambda))| \leq 1$ for all $\lambda\in\RR$. Finally, by integrating it with the smearing functions as described above, we obtain the geodesic decomposition identity \eqref{geodesic_prop}.

\subsection{Transition identity \eqref{transition_id}}
\label{app:transition}

Let us consider the following identity  \cite{Hijano:2015zsa}:
\be 
\label{transition_id_real}
\int_0^\infty\frac{du}{u^2}\int_{\RR}dz\; G_{h_1}(\bx,\bx_1)G_{h_2}(\bx,\bx_2)= \frac{1}{{\dm^{h_1}_{h_2}}}G_{h_1}(\bx_1,\bx_2)+\frac{1}{{\dm^{h_2}_{h_1}}}G_{h_2w}(\bx_1,\bx_2)\,,
\ee
where $z_1,z_2\in\RR$, $u_1,u_2\in\RR_+$ and $\bx_1\neq\bx_2$.  In the decomposition of \ads Feynman diagrams, we encounter  similar integrals of bulk-to-bulk propagators with  endpoints $\bx_1,\bx_2\in\CC\times\CC$. Consequently, the identity \eqref{transition_id_real} must  be analytically continued to the complex plane for  each variable. 

We begin  with the analytic continuation in $z_1$. To this end, consider the following integral
\be 
\label{I}
I_{\RR}(\bx_1,\bx_2;\varepsilon)= \int_{\RR_+\setminus U_{\varepsilon}(u_1)}\frac{du}{u^2}\int_{\RR}dz\; G_{h_1}(\bx,\bx_1)G_{h_2}(\bx,\bx_2)\,,
\ee 
where $U_{\varepsilon}(u_1)$ is the $\varepsilon$-neighbourhood of the point $u_1$. The integrand has four branch cuts, $z=z_i\pm i|u-u_i|t$ with  $t\in[1,\infty)$ for  $i=1,2$; therefore, it is holomorphic for any $z\in\RR$ as long as $|\im(z_1)|<\varepsilon$.

\begin{figure}
\centering
\includegraphics[scale=0.9]{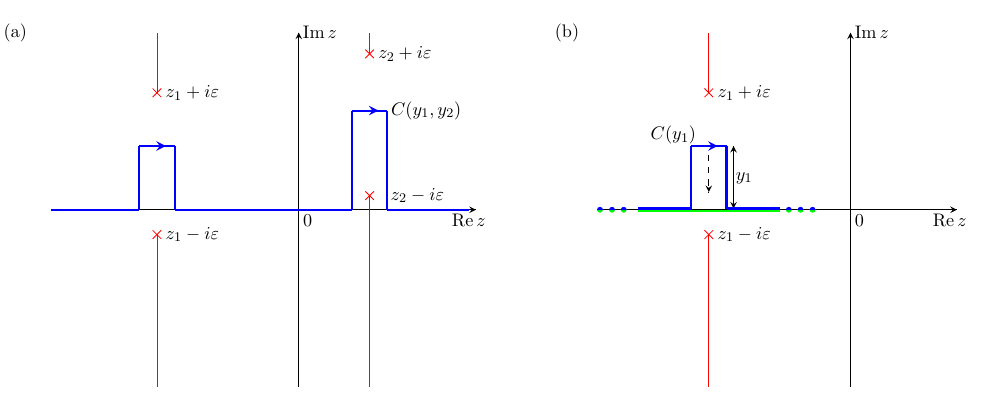}
\caption{{\bf (a)} The contour $C(y_1,y_2)$  on the complex $z$-plane $\mathbb{C}$. {\bf (b)} The blue line corresponds to the contour $C(y_1)$, and the green line corresponds to the straight contour. The dashed arrow indicates the  deformation of $C(y_1)$ into the straight  contour. Note that the restriction $y_1 - \varepsilon<\im(z_i) < \varepsilon$ ensures that no branch points lie on either  contour.}
\label{fig:C_y}
\end{figure}

In order to analytically continue  $I_{\RR}(\bx_1,\bx_2;\varepsilon)$ to the domain $|\im(w_i)|\geq \varepsilon$, we consider the following integral:
\be 
\label{I_mod}
I_{C(y_1)}(\bx_1,\bx_2;\varepsilon)= \int_{\RR_+\setminus U_{\varepsilon}(u_1)}\frac{du}{u^2}\int_{C(y_1)}dz\; G_{h_1}(\bx,\bx_1)G_{h_2}(\bx,\bx_2)\,,
\ee
where $y_i\in\RR$ and $C(y_1) \equiv C(y_1, 0)$, the contour $C(y_1, y_2)$ is shown in fig.~\bref{fig:C_y} {\bf (a)}. The integrand here is a holomorphic function for any $z\in C(y_1)$ only when $|\im(z_1)-y_1|<\varepsilon$. Thus, $I_{C(y_1)}$ \eqref{I_mod} provides the analytic continuation of $I_{\RR}$ \eqref{I} to  $z_1\in\CC$ such that $|\im(z_1)-y_1|< \varepsilon$, where $I_{C(y_1)}$ is holomorphic. By choosing  different values of the  parameters $y_1$, we obtain the analytic continuation for various domains of $z_1$. There are two distinct cases to consider: (1) $|y_1|< 2\varepsilon$ and  (2) $|y_1|\geq2 \varepsilon$.

In the first case $|y_1|< 2\varepsilon$, we fix $y_1 > 0$ without  loss of generality. The case $y_1 < 0$ is analogous,  while  $y_1=0$ is trivial since $C(0) = \RR$ and $I_{C(0)} = I_{\RR}$. To show  that \eqref{I_mod} provides the analytic continuation of  \eqref{I}, we verify  that the  integrals coincide  in their common domain of convergence. The integral $I_{C}$ converges when $y_1 - \varepsilon < \im(z_i) < y_1+\varepsilon$, whereas  the integral $I_{\RR}$ converges when $- \varepsilon < \im(z_i) < \varepsilon$. For  $0<y_1< 2\varepsilon$, these domains  have a non-empty overlap, $y_1 - \varepsilon<\im(z_i) < \varepsilon$. Within this overlap, $I_{\RR}$ and $I_{C}$  coincide because  the contour $C(y_1)$ can be continuously deformed into the real line without crossing any branch cuts of the integrand, see fig.~\bref{fig:C_y} {\bf (b)}. This establishes the analytic continuation for $|y_1|< 2\varepsilon$.

In the second case $y_1\geq2 \varepsilon$ (where we again fix $y_1 > 0$), the domains of convergence for  $I_{C}$ and $I_{\RR}$ do not overlap. To show that one is the analytic continuation of the other, we introduce a sequence of integrals, starting from $I_\RR$ and ending  with $I_{C(y_1)}$, such that only adjacent members have non-empty overlaps of their convergence domains.  This is achieved by defining $n$ auxiliary variables $\alpha_k = ky_1/n$ such that $\alpha_k - \alpha_{k-1} < 2\varepsilon$, which determine a set of  integrals $I_{C(\alpha_k)}$. To show that  $I_{C(\alpha_k)}$  analytically  continues  $I_{C(\alpha_{k-1})}$, we apply  the same arguments as in the case $|y_1|< 2\varepsilon$: the condition $\alpha_k - \alpha_{k-1} < 2\varepsilon$ ensures that their convergence domains intersect.  Within this intersection, the contour $C(\alpha_k)$ can be continuously deformed  into $C(\alpha_{k-1})$ without crossing brunch cuts, as illustrated in fig.~\bref{fig:C_y} {\bf (b)}. Thus, $I_{C(\alpha_k)} = I_{C(\alpha_{k-1})}$ in their common  domain of convergence, which establishes the analytic continuation for $|y_1|\geq 2\varepsilon$.

Note that this proof can also be applied to obtain the analytic continuation of \eqref{I_mod} to $z_2\in\CC$, provided that $\re(z_1)\neq\re(z_2)$. This is achieved by replacing  the contour $C(y_1)$ in \eqref{I_mod} with $C(y_1,y_2)$, see fig.~\bref{fig:C_y} {\bf (a)}: 
\be 
\label{I_z12}
\int_{\RR_+\setminus (U_{\varepsilon}(u_1) \cup U_{\varepsilon}(u_2))}\frac{du}{u^2}\int_{C_z}dz\; G_{h_1}(\bx,\bx_1)G_{h_2}(\bx,\bx_2)\,,
\ee
where $C_z = C\big(\im(z_1),\im(z_2)\big)$. By choosing $y_i=\im(z_i)$, the domain where  \eqref{I_z12} is holomorphic becomes independent of $\varepsilon$: the condition  $|\im(z_i) - y_i|< \varepsilon$ is then satisfied for any $\varepsilon>0$ and $z_i\in\CC$. By taking the limit $\varepsilon\to 0$ we obtain the analytically continued transition identity: 
\be 
\label{transition_z12}
\int_0^\infty\frac{du}{u^2}\int_{C_z}dz\; G_{h_1}(\bx,\bx_1)G_{h_2}(\bx,\bx_2)= \frac{1}{{\dm^{h_1}_{h_2}}}G_{h_1}(\bx_1,\bx_2)+\frac{1}{{\dm^{h_2}_{h_1}}}G_{h_2}(\bx_1,\bx_2)\,,
\ee
where $z_1,z_2\in\CC, \re(z_1)\neq\re(z_2)$, and $u_1,u_2\in\RR_+$. 

The analytic continuation in $u_i$ can be performed  similarly. Since  the contour $C_z$ does not depend on $u$, we may interchange the order of  integration and consider the following integral: 
\be 
\label{I_u}
I^u_{\RR}(\bx_1,\bx_2;\varepsilon) = \int_{C_z\setminus U_{\varepsilon}(z_1)}dz\int_{\RR_+}\frac{du}{u^2}\; G_{h_1}(\bx,\bx_1)G_{h_2}(\bx,\bx_2)\,.
\ee
The integrand has eight branch cuts in the $u$-plane:\footnote{There are four additional  branch cuts arising from the prefactor of the bulk-to-bulk propagators; however,  they do not affect the proof.} $u = \pm u_is\pm i\sqrt{u_i^2(1-s^2)+(z-z_i)^2}$ with $s\in[0,1]$, $i=1,2$. Repeating the procedure described above, we  analytically continue the integral \eqref{I_u} to $u_i\in\CC$ by deforming  the integration contour from $\RR_+$ to $C_u$, as shown in fig.~\bref{fig:C_u}. Note, however, that due to the dependence of the branch cut on the variable $z$, the contour $C_z$ is restricted, see fig.~\bref{fig:C_u}. Taking the limit $\varepsilon\to0$ completes the proof of the transition identity.

\subsection{Integral representations for conformal blocks: a case-by-case study}
\label{app:integral_vertex}

Here, we derive integral representations of the conformal blocks in various domains of conformal weights: $\DD_{34}$, $\DD_{12}\cap\DD_{34}$, and  $\PP_4$. Note that the integral representations  in the domains $\DD_{12}$, $\PP_1$, $\PP_2$, and $\PP_3$  are obtained by relabelling the conformal weights and coordinates as described in subsection \bref{subsec:decom_cases}.

\paragraph{Domain $\DD_{34}$.} Let us consider the following geodesic integral
\be 
\label{alt_geodesic_rep_bndry}
F(z_2,z_3,z_4) = \frac{1}{\beta_{h h_3 h_4}}\int_{\gamma_{34}}d\lambda'\, \cV^{\infty}_{\h_1 \h_2 h}(z_2, \bx(\lambda')) \,K_{h_3}(\bx(\lambda'),z_3)\,K_{h_4}(\bx(\lambda'),z_4)\,,
\ee 
which converges for $\bm h \in \DD_{34}$. $\cV^{\infty}_{\h_1 \h_2 h}(z_2, \bx(\lambda'))$ is the properly rescaled  three-point AdS vertex function with two boundary points, one of which is at infinity  \cite{Alkalaev:2024cje}:
\be 
\label{3pt_vertex_bndry}
\ba{l}
\dps
\cV^{\infty}_{\h_1 \h_2 h}(z_2, \bx) \equiv \lim_{z_1\to\infty} \,\lim_{u_1,u_2\to0}\,
z_1^{2h_1}u_1^{-h_1}u_2^{-h_2}
\cV_{\h_1 \h_2 h}(\bx_1, \bx_2, \bx)
\vspace{2.5mm}  
\\
\dps
= K_{\frac{h+h_2-h_1}{2}}(\bx,z_2)\,{}_2F_1\left[\frac{h-h_1+h_2}{2},\frac{h+h_1-h_2}{2}; h+\half\Big|\,\frac{u^2}{u^2+(z-z_2)^2}\right].
\ea
\ee 
The integral \eqref{alt_geodesic_rep_bndry} represents  the four-point conformal block at the points $(\infty,z_2,z_3,z_4)$ with $\bm h \in \DD_{34}$. To show this, we substitute the AdS vertex function, the bulk-to-boundary propagators \eqref{bulk-to-boundary} and the geodesic \eqref{comp_geod} into the integrand. By changing the variable to $e^{2\lambda} = \frac{t}{1-t}$, applying the Pfaff transformation to the Gauss hypergeometric function in \eqref{3pt_vertex_bndry},
\be
\label{4kummer2f1}
{}_2F_1(a, b; c|z) =(1-z)^{-b}\,{}_2F_1\left(c-a, b; c| \frac{z}{z-1}\right),
\ee
and expanding  the resulting hypergeometric function into a series, we obtain 
\be 
\ba{c}
\dps
F(z_2,z_3,z_4) = \frac{1}{\beta_{h h_3 h_4}} \sum_{n=0}^\infty\, \frac{(-)^n}{2^{2n+1}}\frac{(h+h_2-h_1)_{2n}}{n!(h+\half)_n}\,z_{34}^{h-h_3-h_4+2n}
\vspace{2.5mm}  
\\
\dps
\times\, \int_0^1dt\,t^{\frac{h_4-h_3+h+2n}{2}-1}(1-t)^{\frac{h_3-h_4+h+2n}{2}-1}(z_{24} - z_{34}t)^{h_1-h-h_2-2n}\,,
\ea
\ee 
where $z_{ij} = z_i-z_j$. Integrating over $t$ using the Euler integral representation of the Gauss  hypergeometric function \cite{Bateman:100233}
\be 
\label{hyper_integral}
\dps
{}_2F_1\left(a,b;c\big|z\right) = \sum_{k=0}^\infty\frac{(a)_k(b)_k}{k!(c)_k}\,z^k = \frac{\Gamma(c)}{\Gamma(b)\Gamma(c-b)}\int_0^1 du\; u^{b-1}(1-u)^{c-b-1}\left(1-zu\right)^{-a},
\ee 
yields a double infinite sum:
\be 
\ba{c}
\dps
F(z_2,z_3,z_4) = \frac{1}{\beta_{h h_3 h_4}} \sum_{n,k=0}^\infty\, \frac{(-)^n}{2^{2n+1}}\frac{1}{n!(h+\half)_n}\,\Gamma\Big(\frac{h_3-h_4+h+2n}{2}\Big)
\vspace{2.5mm}  
\\
\dps
\times\, \frac{\Gamma(\frac{h_4-h_3+h+2n}{2}+k)(h+h_2-h_1)_{k+2n}}{k!\Gamma(h+2n+k)}\,
z_{24}^{h_1-h-h_2-2n-k}z_{34}^{h-h_3-h_4+2n+k}\,,
\ea
\ee 
where we have imposed the restriction $|z_{24}|<|z_{34}|$ required for the convergence of the integral. To complete  the proof, we change the summation indices $k = k'-2n$ and  apply  the following identity to the sum over $n$ \cite{Bateman:100233}:
\be 
\sum_{n=0}^{[ \frac{k}{2}+\half ]}\frac{1}{2^{2n}}\frac{(-k)_{2n}(a)_n}{n!(b)_n(a-k-b+\frac{3}{2})_n} = \frac{(2b-2a-1)_k}{(2b-1)_k}\frac{\Gamma(b+k-\half)\Gamma(b-a-\half)}{\Gamma(b-a+k-\half)\Gamma(b-\half)}\,, \quad k\in\NO\,.
\ee 
Introducing the cross-ratio $\frac{z_{12}z_{34}}{z_{24}z_{13}}$, which reduces to $\frac{z_{34}}{z_{24}}$ at the points $\{\infty,z_2,z_3,z_4\}$, we finally obtain: 
\be 
\label{c_block}
\ba{l}
\dps
F(z_2,z_3,z_4) = z_{34}^{h-h_3-h_4}z_{24}^{h_1-h-h_2}{}_2F_1\left(h+h_2-h_1, h-h_3+h_4; 2h\Big| \frac{z_{34}}{z_{24}}\right)
\vspace{2.5mm}  
\\
\dps
\hspace{21mm}= \lim_{z_1\to\infty}z_1^{2h_1}F_{h_1h_2h_3h_4,h}(z_1,z_2,z_3,z_4)\,,
\ea
\ee 
where summing  over $k'$ using \eqref{hyper_integral} has recovered    the Gauss hypergeometric function. The function on the right-hand side of \eqref{c_block} is the global four-point conformal block at the points $\{\infty,z_2,z_3,z_4\}$. We can show that the integral \eqref{alt_geodesic_rep_bndry} is conformally  covariant, i.e. making an $SL(2, \mathbb{R})$  transformation $w(z):\{\infty,z_2,z_3,z_4\}\mapsto \{w_1,w_2,w_3,w_4\}$, we obtain the geodesic representation of the four-point conformal block calculated at arbitrary boundary  points:
\be 
\label{alt_geodesic_rep_bndry_2}
\ba{l}
\dps
F_{h_1h_2h_3h_4,h}(w_1,w_2,w_3,w_4) =  
\vspace{2.5mm}  
\\
\dps
\hspace{10mm}= \frac{1}{\beta_{h h_3 h_4}}\int_{\gamma_{34}}d\lambda'\, \cV_{\h_1 \h_2 h}(w_1, w_2, \bx(\lambda')) K_{h_3}(\bx(\lambda'),w_3)\,K_{h_4}(\bx(\lambda'),w_4)\,.
\ea
\ee 
Substituting this representation into the HKLL reconstruction formula \eqref{HKLL_vertex} completes the proof of \eqref{alt_geodesic_rep}.

\paragraph{Domain $\DD_{12}\cap\DD_{34}$.}   Consider the following representation of a three-point  AdS vertex function with two arbitrary boundary points \cite{Alkalaev:2024cje}:
\be 
\label{3pt_geodesic}
\cV_{ \h_1\h_2 h}(w_1,w_2,\bx) = \frac{1}{\beta_{hh_1h_2}}\int_{\gamma_{12}}d\lambda\,K_{\h_1}(\bx(\lambda), w_1) K_{\h_2}(\bx(\lambda),w_2) G_{h}(\bx(\lambda),\bx)\,,
\ee 
which holds for $h_1 < h + h_2$ and $h_2 < h+h_1$. Substituting this geodesic integral into \eqref{alt_geodesic_rep_bndry_2} yields  the geodesic Witten diagram representation of the four-point conformal block \cite{Hijano:2015zsa}:
\be 
\label{geodesic_rep_bndry}
\ba{c}
\dps
F_{h_1h_2h_3h_4,h}(w_1,w_2,w_3,w_4) = \frac{1}{\beta_{\h \h_1 h_2}\beta_{h h_3 h_4}}\int_{\gamma_{12}}d\lambda \int_{\gamma_{34}}d\lambda' \, K_{\h_1}(\bx(\lambda), w_1)
\vspace{3mm}  
\\
\dps
\times  K_{\h_2}(\bx(\lambda),w_2)\, G_{h}(\bx(\lambda),\bx(\lambda')) K_{h_3}(\bx(\lambda'),w_3)\,K_{h_4}(\bx(\lambda'),w_4)\,.
\ea
\ee 
To obtain the geodesic integral representation of the AdS vertex function \eqref{geodesic_rep}, we substitute  \eqref{geodesic_rep_bndry} into  \eqref{HKLL_vertex}. Note, however, that the  representation \eqref{geodesic_rep_bndry} holds only  when $|z_1-z_2|<|z_i-z_j|$ and $|z_3-z_4|<|z_k-z_l|$ for $i\in\{1,2\}$, $j\in \{3,4\}$ and $k\in \{1,2\}$, $l\in \{3,4\}$; otherwise, the  integration contours may cross the branch cuts of the integrand. 

\paragraph{Domain $\PP_4$.} Let us consider the following integral:
\be 
\label{newF}
F(z_2,z_3,z_4) =  \alpha_{hh_3,n}\oint_{u=0} \frac{du}{u^2} \oint_{P[z_4-iu,z_4+iu]} dz\; \cV^{\infty}_{\h_1 \h_2 h}(z_2, \bx) K_{h_3}(\bx,z_3)\,K_{h_4}(\bx,z_4)\,,
\ee 
which converges for $\bm h \in \PP_4$; $\alpha_{hh_3,n}$ is given by \eqref{alpha_prime}. This integral represents the four-point global conformal block at the points $\{\infty,z_2,z_3,z_4\}$ with  $\bm h \in \PP_4$. To show this, we substitute the AdS vertex function, the bulk-to-boundary propagators and the coefficients into the integral,  change the variable  $z = z_4 + iu(2t-1)$,   apply  the Pfaff transformation \eqref{4kummer2f1} to the Gauss hypergeometric function in \eqref{3pt_vertex_bndry}, and then expand into series to obtain:
\be 
\ba{l}
\dps
F(z_2,z_3,z_4) =  i\alpha_{hh_3,n}\sum_{k=0}^\infty \oint_{u=0} \frac{du}{u^2} \oint_{P[0,1]} dt\; (-)^{k-h_4}2^{-2h_4-2k+1}\frac{(h+h_2-h_1)_{2k}}{k!(h+\half)_k}
\vspace{2.5mm}  
\\
\dps
\times  u^{h_3+h-h_4+2k+1} (t(1-t))^{-h_4}(z_{24}-iu(2t-1))^{h_1-h_2-h-2k}(u^2+(z_{34}-iu(2t-1))^2)^{-h_3}\,.
\ea
\ee
To integrate over $t$ we use the analytically continued integral representation of the Lauricella function \cite{Matsumoto_2020}:
\be 
\label{def_lauricella}
\ba{l}
F^n_\text{D} \left[\begin{array}{cccc}
     a,&b_1, ...\,, &b_n  \\
     &c \end{array};z_1,...\,,z_n\right] =
\vspace{2.5mm}  
\\
\dps
 = \frac{1}{(1-e^{2\pi i a})(1-e^{2\pi i(c-a)})}\frac{\Gamma(c)}{\Gamma(a)\Gamma(c-a)}\oint_{P[0,1]}dx\;x^{a-1}(1-x)^{c-a-1}\prod_{i=1}^n(1-z_ix)^{-b_i}\,,
\ea
\ee 
where $n=3$ in our case and $P[0,1]$ is the Pochhammer contour.  Expanding the resulting Lauricella function into a series yields:
\be 
\ba{l}
\dps
F(z_2,z_3,z_4) =  -i\frac{(-)^{n}}{\pi^{\frac{1}{2}}}\frac{\Gamma(2h_4-1)n!}{\Gamma(h_4)\Gamma(h+h_3+n-\half)(h)_n(h_3)_n}\sum_{k,s_1,s_2,s_3=0}^\infty \oint_{u=0} \frac{du}{u^2}  (-)^{k}
\vspace{2.5mm}  
\\
\dps
2^{-2h_4-2k+s_1+s_2+s_3+1}\, i^{s_1+s_2+s_3} \frac{(1-h_4)_{s_1+s_2+s_3}(h_3)_{s_1}(h_3)_{s_2}(h_2+h-h_1)_{s_3+2k}}{k!(h+\half)_ks_1!s_2!s_3!(2-2h_4)_{s_1+s_2+s_3}}\,
\vspace{2.5mm}  
\\
\dps
\times\, u^{h_3+h-h_4+2k+s_1+s_2+s_3+1}(z_{24}+iu)^{h_1-h_2-h-2k-s_3}z_{34}^{-h_3-s_1}(z_{34}+2iu)^{-h_3-s_2}\,.
\ea
\ee
To integrate over $u$, we note that due to the restriction $\bm h \in \PP_4$, the power of $u$ in the integrand is an integer; therefore,  the residue theorem can be applied to the integral over $u$. Summing the result over $k$ using the following identity \cite{Bateman:100233}: 
\be 
\label{2F1_in_1}
\sum_{k=0}\frac{(a)_k(b)_k}{k!(c)_k} = \frac{\Gamma(c)\Gamma(c-a-b)}{\Gamma(c-a)\Gamma(c-b)}\,,
\qquad \text{Re}(c-a-b)>0\,,
\ee 
we obtain: 
\be 
\ba{l}
\dps
F(z_2,z_3,z_4) =  \frac{2(-)^{n}\pi^{\frac{1}{2}}\Gamma(2h_4-1)n!}{\Gamma(h_4)\Gamma(h+h_3+n-\half)(h)_n(h_3)_n}\sum_{m,s_1,s_2,s_3=0}^\infty  (-)^{n+s_1+s_2+s_3}2^{2n-2h_4+1} 
\vspace{2.5mm}  
\\
\dps
\frac{(h)_{2n-s_1-s_2-s_3-m}}{m!(2n-s_1-s_2-s_3-m)!}\frac{(1-h_4)_{s_1+s_2+s_3}(h_3)_{s_1}(h_3)_{s_2+m}(h_2+h-h_1)_{2n-s_1-s_2-m}}{s_1!s_2!s_3!(2-2h_4)_{s_1+s_2+s_3}(2h)_{2n-s_1-s_2-s_3-m}}\,
\vspace{2.5mm}  
\\
\dps
\times\, z_{24}^{h_1-h_2-h-2n+s_1+s_2+m}z_{34}^{-2h_3-s_1-s_2-m}\,,
\ea
\ee
where we also used $h_4 = h_3+h+2n$ to make the result more compact. Next, we redefine  the summation indices  as $s_2=s_2'-s_1-m$, sum over $s_1$ using \eqref{2F1_in_1}, and take into account the poles of the gamma function to restrict the domain of $m$.  Finally, applying  the following identity to the sum over $m$ \cite{prudnikov1986integrals}:
\be 
\sum_{k=0}^{n}\frac{(-n)_k(a)_k(b)_k}{k!(c)_k(d)_k} = \frac{(d-b)_n}{(d)_n}\sum_{k=0}^{n}\frac{(-n)_k(c-a)_k(b)_k}{k!(c)_k(b-n-d+1)_k}\,,
\ee 
we find 
\be 
\label{interm_spec}
\ba{l}
\dps
F(z_2,z_3,z_4) =  \frac{2(-)^{n}\pi^{\frac{1}{2}}\Gamma(2h_4-1)n!}{\Gamma(h_4)\Gamma(h+h_3+n-\half)(h)_n(h_3)_n}\sum_{m,s'_2,s_3=0}^\infty  (-)^{n-h_4+s_3+m}2^{2n-2h_4+1} 
\vspace{2.5mm}  
\\
\dps
\frac{\Gamma(2h+4n-s'_2-s_3)(h)_{2n+m-s'_2-s_3}}{m!(s'_2-m)!\Gamma(2h+4n-2s'_2-s_3+m)}\frac{(1-h_4)_{s'_2+s_3-m}(h_2+h-h_1)_{2n-s'_2}}{(2n-s'_2-s_3)!s_3!(2-2h_4)_{s'_2+s_3-m}(2h)_{2n-s'_2-s_3}}\,
\vspace{2.5mm}  
\\
\dps
\times  z_{24}^{h_1-h_2-h-2n+s'_2}z_{34}^{-2h_3-s'_2}\,.
\ea
\ee
By changing the summation indices  $m= m'+s_3$ and  $s'_2=2n-k$, restricting the  domains of $m'$ and $s_3$ according to the poles of the gamma function, and performing the summations over $s_3$ and then over $m'$ using the identities \cite{prudnikov1986integrals}: 
\be 
\ba{c}
\dps
\sum_{k=\max(0,-m)}^n\frac{(-n)_k(b)_k(c)_k}{k!(a+b+c-d+1)_k(m)_k} = \frac{(m-a)_n(m-b)_n}{(m)_n(m-a-b)_n}\,,
\quad n\in\NN_{0}\,,\;m\in\ZZ\,,
\vspace{2.5mm}  
\\
\dps
\sum_{k=0}^{2n}\frac{(-2n)_k(b)_k(c)_k}{k!(2b)_k(\frac{c-2n+1}{2})_k} = \frac{(\half)_n(b+\frac{1-c}{2})_n}{(\frac{1-c}{2})_n(b+\half)_n}\,,
\quad n\in\NN_{0}\,,
\ea
\ee 
we find that 
\be 
\ba{l}
\dps
F(z_2,z_3,z_4) = z_{34}^{h-h_3-h_4}z_{24}^{h_1-h-h_2}{}_2F_1\left(h+h_2-h_1, h-h_3+h_4; 2h\Big| \frac{z_{34}}{z_{24}}\right)
\vspace{2.5mm}  
\\
\dps
\hspace{21mm}= \lim_{z_1\to\infty}z_1^{2h_1}F_{h_1h_2h_3h_4,h}(z_1,z_2,z_3,z_4)\,.
\ea
\ee 
Here, the sum over $n$ has been performed to recover the Gauss hypergeometric function \eqref{hyper_integral}, followed by the Pfaff transformation \eqref{4kummer2f1}. The result reproduces  the global conformal block  with conformal weights  constrained as $h_4 = h+h_3+2n$ at the points $\{\infty,z_2,z_3,z_4\}$. By applying a conformal transformation $w(z):\{\infty,z_2,z_3,z_4\} \mapsto \{w_1,w_2,w_3,w_4\}$ and using the conformal covariance of \eqref{newF}, we obtain the representation \eqref{non_triangle_rep}.

\section{Regularization procedure}
\label{app:regularization}

The expression  \eqref{new_aleph} and the first five lines of the expression \eqref{new_beth}  represent the coefficients of the single-trace terms in the expansions of the four-point contact \eqref{vert_decomp} and exchange \eqref{exch_final} AdS Feynman diagrams, respectively. These coefficients along with  the  single-trace terms are produced by the first terms of \eqref{cont_mod} and \eqref{ex_start} upon applying  the decomposition algorithm in section \bref{sec:algorithm}. However, a naive application of step II of the algorithm, which involves interchanging  integrations and summations, leads to divergent coefficients. One of the sources of this divergence is the integral over $\bx$, since its integration domain includes the points where the integrand develops  a logarithmic singularity. The integrand remains  integrable, but the divergence may become problematic after interchanging the order of the integration over $\bx$ and other limiting operations. Note that the integration domains for the integrals over $\bx$ in the second and third terms of \eqref{contact_interm_1} and in the last term of \eqref{exch_int} are compact and do not include the singularities of the integrand; therefore, no divergences arise during the decomposition process. Also note that the integration domains for the  integrals over $\bx$ in the second and third terms in \eqref{exch_int} are non-compact and contain singularities of the integrand; therefore, these terms should be regularized in the same manner as the first terms of \eqref{contact_interm_1} and \eqref{exch_int}. We omit the procedure of interchanging limiting operations for these terms because  it is almost identical to the procedure for the first terms (the resulting regularized coefficients are listed in \eqref{new_beth}).

In this Appendix, we demonstrate how to regularize the resulting  divergences and rigorously interchange the limiting operations by introducing a regularization factor. For the reader's convenience, we recall a few  standard theorems from mathematical analysis required for our consideration, presenting them as they appear throughout the discussion.

The regularization procedure  consists of three steps: (1) considering the first terms of \eqref{cont_mod} and \eqref{ex_start}; (2) applying   the geodesic decomposition identity to them, following step II of the decomposition algorithm in section \bref{sec:algorithm}; (3) introducing  a regularization factor and interchanging  the corresponding integrations and summations. This procedure can be simplified by making use of the following observations about the interplay between the multiple integrals, their integrands, and sums appearing in these terms: 
\begin{enumerate}
    \item When deriving  \eqref{contact_interm_1} and \eqref{exch_int} from respectively \eqref{cont_mod} and \eqref{ex_start}, we  do not interchange the integrations over $w_i$ with other limiting operations.
    \item The boundary values of the first terms of \eqref{cont_mod} and \eqref{ex_start}  can be obtained from the latter by replacing $\widehat{G}_h(\bx,\bx_1,w)$ with $ K_h(\bx,z_1)$, removing integrations over $w_i$ and changing the complex integration contours over $\bx,\bx'$ to the standard AdS$_2$ half-plane, $\bx,\bx'\in\RR_+\times\RR$.
    \item The modified propagator $\widehat{G}_{h_i}(\bx,\bx_i,w_i)$ is the product of the bulk-to-boundary propagator $K_{h_i}(\bx,w_i)$ and the smearing function $\mathbb{K}_{h_i}(\bx_i,w_i)$. When interchanging the limiting operations one can treat the smearing functions inside the modified propagators in  the first terms of \eqref{cont_mod} and \eqref{ex_start} as constants; to prove this one uses the observation 1 and the fact that $\bx_i$ are the arguments of the AdS Feynman diagrams, i.e. they are fixed.
    \item The boundary geodesic decomposition identity \eqref{bndry_geodesic_prop} should be applied to the product of two bulk-to-boundary propagators $K$ instead of the product of two modified propagators $\widehat{G}$. The result of the application of this identity is similar to the bulk version \eqref{geodesic_prop}, with the modified propagators $\widehat{G}$ changed to bulk-to-boundary propagators $K$.
    \item The first terms of \eqref{cont_mod} and \eqref{ex_start} can be reconstructed from their boundary values using the HKLL construction.
\end{enumerate}

Therefore, instead of regularizing functions in the bulk of AdS, one can regularize their boundary values. This involves applying  to them  the boundary geodesic decomposition identity \eqref{bndry_geodesic_prop} as required by step II of the decomposition algorithm, introducing a regularization factor, interchanging the order of the limiting operations,  and then uplifting  the resulting functions back to the bulk. Observations 2, 3, and 4 ensure that the geodesic decomposition identity can be applied to the boundary values of the first terms of \eqref{cont_mod} and \eqref{ex_start} and that the resulting expression  is structurally the same, up to the changes described in observation 2. Observation 1 ensures that the absence  of the integrations over the boundary variables $w_i$ does not affect the procedure of interchanging the limiting operations, while the observation 5 ensures that the original bulk function can be reconstructed from the boundary. The simplification occurs due to the change in the contours in observation 2 and the change of the argument of the bulk-to-boundary propagator $K_{h_i}(\bx,w_i)$  from the complex $w_i$ to the real $z_i$, as seen from observations 2 and 3.

The boundary values of the first terms \eqref{cont_mod} and \eqref{ex_start} are the contact and exchange four-point Witten diagram \eqref{4pt_contact_expansion} and  \eqref{4pt_exchange_expansion}, respectively. Applying the boundary geodesic decomposition identity to the bulk-to-boundary propagators $K_{\h_1}(\bx,z_1) K_{\h_2}(\bx,z_2)$ in the Witten diagrams in the presence of  a regularization factor, we obtain
\be
\label{start_reg}
\ba{l}
\dps
F^{\text{c/e}}(z_1,z_2) = \int_{\text{AdS}_2}d^2\bx\sqrt{g(\bx)} \,\sum_{n=0}^{\infty}\, \lim_{\varepsilon\to0^+}(1-\varepsilon)^n\frac{a^{h_1h_2}_n}{\beta_{h_{12|n}h_1h_2}}
\vspace{2.5mm}  
\\
\dps
\hspace{24mm}\times \int_{\gamma_{12}}d\lambda\;K_{h_1}(\bx(\lambda),z_1)K_{h_2}(\bx(\lambda),z_2)G_{h_{12|n}}(\bx(\lambda),\bx)R^{\,\text{c/e}}(\bx)\,,
\ea
\ee 
where the $R$-functions are given by
\be 
\ba{l}
\dps
R^{\,\text{c}}(\bx) = K_{h_3}(\bx,z_3)K_{h_4}(\bx,z_4)\,,
\vspace{2.5mm}  
\\
\dps
R^{\,\text{e}}(\bx) = \int_{\text{AdS}_2}d\bx\,G_{\tilde{h}}(\bx,\bx')K_{h_3}(\bx',z_3)K_{h_4}(\bx',z_4)\,.
\ea
\ee 
Note that we did not apply the geodesic decomposition identity to the bulk-to-boundary propagators $K_{\h_3}(\bx,z_3)K_{\h_4}(\bx,z_4)$ in the $R$-functions to simplify the analysis. Instead, this identity is  applied after the interchange of the limiting operations in \eqref{start_reg} at the end of this Appendix.

To interchange the integration over AdS$_2$ with the other limiting operations in \eqref{start_reg} we make several interchanges in the following order:
\begin{enumerate}
    \item Interchanging the limit $\varepsilon\to 0^+$ and the infinite sum over $n$. 
    \item Interchanging the limit $\varepsilon\to 0^+$ and the integral over AdS$_2$.
    \item Interchanging the infinite sum over $n$ and the integral over AdS$_2$.
    \item Interchanging the integral over the geodesic $\gamma_{12}$ and the integral over AdS$_2$.
\end{enumerate}

\paragraph{Step 1.} To interchange the limit and the infinite sum, we  employ the Moore–Osgood theorem \cite{taylor2012general}:

{\it Let $\{a_k(\varepsilon), k\in \NN\}$ be a sequence of functions on a space $X$. If $\dps \lim_{k\to\infty}a_k(\varepsilon) = b(\varepsilon)$ uniformly with respect to $\varepsilon\in X$ and $\dps \lim_{\varepsilon\to0^+}a_k(\varepsilon) = c_k$ for each large $k$, then $\dps \lim_{\varepsilon\to0^+}b(\varepsilon) = \lim_{k\to\infty}c_k < \infty$.} 

To apply the Moore--Osgood theorem to \eqref{start_reg}, we rewrite the function $F^{\text{c/e}}(z_1,z_2)$ as 
\be 
\label{reg_moore_osgood}
F^{\text{c/e}}(z_1,z_2) = \int_{\text{AdS}_2}d^2\bx\sqrt{g(\bx)} \, R^{\,\text{c/e}}(\bx)\lim_{k\to\infty}\,\lim_{\varepsilon\to0^+}\,a_k(\varepsilon)\,,
\ee 
and consider the  sequence of functions 
\be 
\label{moore-osgood}
a_k(\varepsilon) = \sum_{n=0}^k s_n(\varepsilon)\,,
\ee 
where
\be
\label{s_n}
\ba{l}
s_n(\varepsilon) = a^{h_1h_2}_n (1-\varepsilon)^nK_{h_1+n}(\bx,z_1)K_{h_2+n}(\bx,z_2)z_{12}^{2n}
\vspace{2.5mm}  
\\
\dps
\hspace{12mm}
\times\,{}_2F_1\left(h_1+n, h_2+n; h_1+h_2+2n+\half\Big| \frac{u^2z_{12}^2}{(u^2+(z_1-z)^2)(u^2+(z_2-z)^2)}\right).
\ea
\ee 
The terms $s_n(\varepsilon)$ are obtained from \eqref{start_reg} by taking the integrand over $\bx$ and the summand over $n$ without the $R$-function, and then integrating over $\lambda$ using \eqref{3pt_geodesic}. To prove that the sum $a_k(\varepsilon)$ converges uniformly in $\varepsilon$, we use the Weierstrass M-test. To this end, we note that by construction the argument of the Gauss hypergeometric function in \eqref{s_n} satisfies  the following inequality:
\be 
\label{X_def}
0\leq X\leq 1\,, 
\quad 
\bx\in\text{AdS}_2\,, 
\;
z_1,z_2\in\RR\,, 
\quad \text{where}\;
X = \frac{u^2z_{12}^2}{(u^2+(z_1-z)^2)(u^2+(z_2-z)^2)}\,.
\ee 
Since ${}_2F_1(a,b,c|X)$ with $a,b,c > 0$\footnote{In our case, $a=h_1+n>0$, $b=h_2+n>0$,  and $c=h_1+h_2+2n+\half>0$ due to the restriction on the weights $h_i\geq\half$ \eqref{half_restrictions}.} and $X\in[0,1]$ is a monotonically increasing function with its maximum value  given by ${}_2F_1(a,b,c|1) = \frac{\Gamma(c)\Gamma(c-b-a)}{\Gamma(c-a)\Gamma(c-b)}$, one estimates an individual  term $s_n(\varepsilon)$ in \eqref{moore-osgood} as
\be 
\label{a_estimate}
|s_n(\varepsilon)| \leq  K_{h_1+n}(\bx,z_1)K_{h_2+n}(\bx,z_2)z_{12}^{2n}
\frac{(\h_1)_n(\h_2)_n}{n!(\h_1+\h_2-\half+n)_n}\frac{\Gamma(h_1+h_2+2n+\half)\Gamma(\half)}{\Gamma(h_1+n + \half)\Gamma(h_2+n+\half)}\,,
\ee 
where we substituted the coefficient $a^{h_1h_2}_n$ \eqref{a_beta} into \eqref{s_n}. The infinite sum over $n$ of the estimates on the right-hand side of \eqref{a_estimate} converges absolutely as it is the hypergeometric series ${}_5F_4[..., X]$, with $X < 1$ for almost every $\bx\in\text{AdS}_2$ and for all $z_1,z_2\in\RR$. The set of points $\bx$ in which $X = 1$ is a set of measure zero in AdS$_2$, which can  therefore be removed from the integration domain in \eqref{start_reg}. As the resulting majorant hypergeometric function does not depend on $\varepsilon$ and $k$, the sequence $a_k(\varepsilon)$ converges uniformly in $\varepsilon$ by the Weierstrass M-test and the limit $\varepsilon\to0^+$ exists for any $k$. This allows one to apply the Moore-Osgood theorem to interchange the infinite summation and the limit $\varepsilon\to0^+$ in \eqref{start_reg}, producing
\be
\label{2_reg}
\ba{l}
\dps
F^{\text{c/e}}(z_1,z_2) = \int_{\text{AdS}_2}d^2\bx\sqrt{g(\bx)} \,\lim_{\varepsilon\to0^+}\sum_{n=0}^{\infty}\, (1-\varepsilon)^n\frac{a^{h_1h_2}_n}{\beta_{h_{12|n}h_1h_2}}
\vspace{2.5mm}  
\\
\dps
\hspace{24mm}\times \int_{\gamma_{12}}d\lambda\;K_{h_1}(\bx(\lambda),z_1)K_{h_2}(\bx(\lambda),z_2)G_{h_{12|n}}(\bx(\lambda),\bx)R^{\,\text{c/e}}(\bx)\,.
\ea
\ee 

\paragraph{Step 2.} To interchange the limit $\varepsilon\to0^+$ and the integral over AdS$_2$ in \eqref{2_reg}, we employ  Lebesgue's dominated convergence theorem \cite{taylor2012general}: 

{\it Let \{$f_k(x),\, k \in \NN\}$  be a sequence of measurable functions on a space $X$ with the measure $
\mu$ such that $\dps\lim_{k\to\infty} f_k(x)= f(x)$. Assume that there exists an integrable function $g(x)$ such that $|f_k(x)|\leq g(x)$ for any  $k\in \NN$. Then, 
\be 
\lim_{k\to\infty} \int_X  f_k(x)d\mu= \int_X f(x)d\mu \,.
\ee} 

In our case, in order to construct the sequence of functions $f_k(x)$ we replace the limit $\varepsilon\to0^+$ in \eqref{start_reg} with $\varepsilon_k\to0^+$ as $k\to\infty$, where $\varepsilon_k = \frac{1}{k+1}$. Then, the functions $f_k(x)$ are given by
\be
\label{fk} 
\ba{l}
\dps
f_k(\bx) = R^{\,\text{c/e}}(\bx)\sum_{n=0}^\infty (1-\varepsilon_k)^n\, a^{h_1h_2}_n\, K_{h_1+n}(\bx,z_1)K_{h_2+n}(\bx,z_2)z_{12}^{2n}
\vspace{2.5mm}  
\\
\dps
\hspace{14mm}
\times \, {}_2F_1\left(h_1+n, h_2+n; h_1+h_2+2n+\half\Big| \frac{u^2z_{12}^2}{(u^2+(z_1-z)^2)(u^2+(z_2-z)^2)}\right),
\ea
\ee 
which are obtained from \eqref{2_reg} by taking the integrand over $\bx$ and integrating over $\lambda$ using \eqref{3pt_geodesic}. To find an appropriate estimate for these functions, we represent  the Gauss hypergeometric function in \eqref{fk} as  a power series and expand the factor $(1-\varepsilon_k)^n$ using the binomial theorem:
\be 
\ba{l}
\dps
f_k(\bx) = R^{\,\text{c/e}}(\bx)\sum_{n=0}^\infty\sum_{l=0}^n\sum_{p = 0}^\infty \,\frac{(-)^n\, \varepsilon_k^l}{l!(n-l)!p!}\,\frac{(h_1)_{n+p}(h_2)_{n+p}}{(h_1+h_2-\half+n)_n(h_1+h_2+\half+2n)_p}
\vspace{2.5mm}  
\\
\dps
\hspace{14mm}
\times \, \Big(\frac{u}{u^2+(z_1-z)^2}\Big)^{h_1+n+p}\Big(\frac{u}{u^2+(z_2-z)^2}\Big)^{h_2+n+p}z_{12}^{2n+2p}\,.
\ea
\ee 
Making the substitution  $n' = p+n$ and summing over $p$ using the following identity \cite{Bateman:100233}
\be 
\sum_{k=0}^\infty\frac{(-l)_k(\frac{a}{2}+1)_k(a)_k}{k!(\frac{a}{2})_k(a+n+1)_k} = \frac{(n-l)(n+1)_{l-1}}{(a+n+1)_{l}}\,,
\ee
we obtain
\be 
\ba{l}
\dps
f_k(\bx) = K_{h_1}(\bx,z_1)K_{h_2}(\bx,z_2)R^{\,\text{c/e}}(\bx) + R^{\,\text{c/e}}(\bx)\sum_{l=1}^\infty\sum_{n=l}^\infty(-\varepsilon_k)^l\, \frac{(h_1)_{n'}(h_2)_{n'}\Gamma(2n'-l)}{(l-1)!(n'-l)!n'!}
\vspace{2.5mm}  
\\
\dps
\hspace{14mm}
\times\,\frac{\Gamma(h_1+h_2-\half+l)}{\Gamma(h_1+h_2-\half+2n')}\Big(\frac{u}{u^2+(z_1-z)^2}\Big)^{h_1+n'}\Big(\frac{u}{u^2+(z_2-z)^2}\Big)^{h_2+n'}z_{12}^{2n'}\,.
\ea
\ee 
To find an estimate for the absolute value of $f_k(\bx)$, we use  the following relation for the above summations  over $l$ and $n$,
\be 
\bigg|\sum_{k=0}^\infty a_k x^k\bigg| \leq \sum_{k=0}^\infty |a_k||x^k|\,, 
\ee 
and note that 
\be  
\bigg|\frac{1}{3}\frac{2^{2n'}(h_1)_{n'}(h_2)_{n'}}{(h_1+h_2-\half)_{2n'}}\bigg|\leq 1, \quad \forall n'\in\NN\,,
\ee 
which allows us to write
\be 
\ba{l}
\dps
|f_k(\bx)| \leq K_{h_1}(\bx,z_1)K_{h_2}(\bx,z_2)R^{\,\text{c/e}}(\bx) + 3K_{h_1}(\bx,z_1)K_{h_2}(\bx,z_2)R^{\,\text{c/e}}(\bx)\Big(\frac{1}{(1-\varepsilon_k)^{h_1+h_2-\half}} - 1\Big) 
\vspace{2.5mm}  
\\
\dps
\hspace{12mm}
\leq 4 K_{h_1}(\bx,z_1)K_{h_2}(\bx,z_2)R^{\,\text{c/e}}(\bx)\,, \qquad k \in \NN\,,
\ea
\ee 
where we used $|X|\leq 1$, summed over $n'$ using 
\be 
\sum_{n=k}^\infty\frac{\Gamma(2n-k)}{n!(k-1)!(n-k)!}2^{-2n} = \frac{1}{2^kk!}\,,
\ee 
and finally summed over $k$ by means  of  the binomial theorem. The resulting bound function is integrable, since it coincides with the integrand of the four-point contact/exchange Witten diagram, both of which are finite.\footnote{These Witten diagrams converge provided that the conformal weights satisfy  the triangle inequalities \eqref{triangle_identity1}, \eqref{triangle_identity2}. In such a case, the values of these integrals are given by \eqref{4pt_contact_expansion}, \eqref{4pt_exchange_expansion}.} It also follows that the functions $f_k(\bx)$ are measurable. Therefore, Lebesgue's theorem  can be applied to interchange the limit $\varepsilon\to0^+$ and the integral over $\bx$ in \eqref{2_reg}: 
\be
\label{3_reg}
\ba{l}
\dps
F^{\text{c/e}}(z_1,z_2) = \lim_{\varepsilon\to0^+}\int_{\text{AdS}_2}d^2\bx\sqrt{g(\bx)} \,\sum_{n=0}^{\infty}\, (1-\varepsilon)^n\frac{a^{h_1h_2}_n}{\beta_{h_{12|n}h_1h_2}}
\vspace{2.5mm}  
\\
\dps
\hspace{24mm}\times \int_{\gamma_{12}}d\lambda\;K_{h_1}(\bx(\lambda),z_1)K_{h_2}(\bx(\lambda),z_2)G_{h_{12|n}}(\bx(\lambda),\bx)R^{\,\text{c/e}}(\bx)\,.
\ea
\ee 

\paragraph{Step 3.} To interchange the infinite sum over $n$ and the integral over AdS$_2$ in \eqref{3_reg} we again apply Lebesgue's theorem. The sequence of functions is given by 
\be 
\label{step_3_f}
\ba{l}
\dps
\tilde{f}_k(\bx) = R^{\,\text{c/e}}(\bx)\, a_k(\varepsilon) = R^{\,\text{c/e}}(\bx)\sum_{n=0}^k a^{h_1h_2}_n (1-\varepsilon)^nK_{h_1+n}(\bx,z_1)K_{h_2+n}(\bx,z_2)z_{12}^{2n}
\vspace{2.5mm}  
\\
\dps
\hspace{14mm}
\times\,{}_2F_1\left(h_1+n, h_2+n; h_1+h_2+2n+\half\Big| \frac{u^2z_{12}^2}{(u^2+(z_1-z)^2)(u^2+(z_2-z)^2)}\right),
\ea
\ee 
where $a_k(\varepsilon)$ is defined in \eqref{moore-osgood}. {The functions $f_k(\bx)$ are measurable because  the product of bulk-to-boundary propagators $K_h(\bx,z)$ and the $R$-function is measurable, as noted earlier, while the Gauss hypergeometric function in \eqref{step_3_f} is  bounded for all $|X|\leq1$.} Using the estimate for the absolute value of the terms $s_n(\varepsilon)$  \eqref{a_estimate} we obtain
\be 
\label{step_3}
\ba{l}
\dps
|\tilde{f}_k(\bx)| \leq  K_{h_1}(\bx,z_1)K_{h_2}(\bx,z_2)R^{\,\text{c/e}}(\bx)
\vspace{2.5mm}  
\\
\dps
\hspace{16mm}
\times\,\sum_{n=0}^\infty \frac{(\h_1)_n(\h_2)_n}{n!(\h_1+\h_2-\half+n)_n} \frac{\Gamma(h_1+h_2+2n+\half)\Gamma(\half)}{\Gamma(h_1+n + \half)\Gamma(h_2+n+\half)} (1-\varepsilon)^n\,,
\ea
\ee 
where we took into account that the function $R^{\,\text{c/e}}(\bx)$ is non-negative and used $|X|\leq 1$. The resulting sum on the right-hand side of \eqref{step_3} is a hypergeometric series ${}_5F_4[...,1-\varepsilon]$, which converges for all  $\varepsilon \in (0, 2)$. As this sum does not depend on $\bx$, the integrability of the estimate is regulated by the prefactor, which is integrable as discussed above. This allows one to apply Lebesgue's theorem to interchange the infinite sum over $n$ and the integral over AdS$_2$. 

\paragraph{Step 4.} Finally, we interchange the order of the integrals over $\bx$ and $\lambda$. This can be done using Tonelli's theorem: 

{\it
If $f(x,y)$ is a non-negative measurable function on $X\times Y$, then 
\be 
\int_{Y}\Big(\int_{X}f(x,y)d\mu(x)\Big)d\nu(y)= \int_{X}\Big(\int_{Y}f(x,y)d\nu(y)\Big)d\mu(x) = \int_{X\times Y}f(x,y)d(\mu\times\nu)\,,
\ee 
where $\mu(x)$ and $\nu(y)$ are measures on $X$ and $Y$, respectively.
}

 As the integrand is non-negative and measurable with respect to $\bx\in \text{AdS}_2$ and $\lambda\in\RR$, the result is given by
\be
\label{4_reg}
\ba{l}
\dps
F^{\text{c/e}}(z_1,z_2) = \lim_{\varepsilon\to0^+}\sum_{n=0}^{\infty}\, (1-\varepsilon)^n\frac{a^{h_1h_2}_n}{\beta_{h_{12|n}h_1h_2}}\int_{\gamma_{12}}d\lambda
\vspace{2.5mm}  
\\
\dps
\hspace{24mm}\times \int_{\text{AdS}_2}d^2\bx\sqrt{g(\bx)} \;K_{h_1}(\bx(\lambda),z_1)K_{h_2}(\bx(\lambda),z_2)G_{h_{12|n}}(\bx(\lambda),\bx)R^{\,\text{c/e}}(\bx)\,,
\ea
\ee 
which is desired integral \eqref{start_reg} with interchanged order of integrations and summations.

To obtain the regularized first term of \eqref{contact_interm_1} or \eqref{exch_int}, we first apply the boundary version of the geodesic decomposition identity \eqref{bndry_geodesic_prop} to the bulk-to-boundary propagators in the function $R^{\,\text{c/e}}(\bx)$. This produces the expression similar to \eqref{start_reg} with a different $R$-function. Then, we interchange the order of geodesic integrals and infinite sums produced by the geodesic decomposition identity with the integral over $\bx$ in the same manner as above. For example, in the case of the contact diagram the result is given by
\be
\label{4_reg_2}
\ba{l}
\dps
F^{\text{c}}(z_1,z_2,z_3,z_4) = \lim_{\varepsilon_1\to0^+}\sum_{n=0}^{\infty}\, (1-\varepsilon_1)^n\frac{a^{h_1h_2}_n}{\beta_{h_{12|n}h_1h_2}}\int_{\gamma_{12}}d\lambda
\lim_{\varepsilon_2\to0^+}\sum_{m=0}^{\infty}\, (1-\varepsilon_2)^n\frac{a^{h_3h_4}_m}{\beta_{h_{34|m}h_3h_4}}\int_{\gamma_{34}}d\lambda'
\vspace{2.5mm}  
\\
\dps
\hspace{31mm}\times \int_{\text{AdS}_2}d^2\bx\sqrt{g(\bx)} \;K_{h_1}(\bx(\lambda),z_1)K_{h_2}(\bx(\lambda),z_2)G_{h_{12|n}}(\bx(\lambda),\bx)
\vspace{2.5mm}  
\\
\dps
\hspace{31mm}\times\, G_{h_{34|m}}(\bx(\lambda'),\bx)K_{h_3}(\bx(\lambda'),z_3)K_{h_4}(\bx(\lambda'),z_4)\,.
\ea
\ee 
Analytically continuing the result to complex $z_i$ by deforming integration contours in $z$-plane and $u$-plane to $\widetilde{C}$ and $\widetilde{C}_u$ and integrating the resulting analytically continued function with the smearing functions over the complex boundary we obtain the first term in \eqref{contact_interm_1} or \eqref{exch_int} (note that the regularization factor there is omitted). After applying steps III and IV of the decomposition algorithm to the first term in \eqref{contact_interm_1} or \eqref{exch_int}, the regularization factor $(1-\varepsilon)^n$ appears in the coefficients of the expansion, which resolves the apparent divergence.

\bibliographystyle{JHEP}
\bibliography{refs}

\end{document}